\documentclass[a4paper,fleqn,usenatbib]{mnras}

\usepackage{newtxtext,newtxmath}
\usepackage[T1]{fontenc}
\usepackage{ae,aecompl}

\usepackage{graphicx}	% Including figure files
\usepackage{amsmath}	% Advanced maths commands
\usepackage{amssymb}	% Extra maths symbols
\usepackage{adjustbox}
\usepackage{xspace}
\usepackage{natbib}
\setcitestyle{notesep={}}
\usepackage{etoolbox}
\makeatletter
\patchcmd\@combinedblfloats{\box\@outputbox}{\unvbox\@outputbox}{}{%
  \errmessage{\noexpand\@combinedblfloats could not be patched}%
}%
\makeatother
\usepackage{cleveref}
\crefname{figure}{Figure}{Figures}
\crefname{section}{Section}{Sections}
\crefname{table}{Table}{Tables}

\newcommand{\Msol}{M$_{\odot}$\xspace}

\newcommand{\squotes}[1]{\lq {#1}\rq\xspace}
\newcommand{\dquotes}[1]{\lq\lq {#1}\rq\rq\xspace}
\newcommand{\eagle}{{\sc eagle}\xspace}

\newcommand{\subfind}{{\sc subfind}\xspace}

\newcommand{\Lbol}{$L_{\mathrm{bol}}$\xspace}

\newcommand{\M}[1]{$M_{\mathrm{#1}}$\xspace}

\newcommand{\edd}{$\lambda_{\mathrm{edd}}$\xspace}
\newcommand{\ndyn}{$n_{\mathrm{dyn}}$\xspace}
\newcommand{\ndynbar}{$|n_{\mathrm{dyn}}|$\xspace}
\newcommand{\ndynbarmajor}{$|n_{\mathrm{dyn[Major]}}|$\xspace}

\newcommand{\erg}{erg~s$^{-1}$\xspace}

\newcommand{\ndynmajor}{$n_{\mathrm{dyn[Major]}}$\xspace}
\newcommand{\ndynmajorbar}{$|n_{\mathrm{dyn[Major]}}|$\xspace}
\newcommand{\ndynminor}{$n_{\mathrm{dyn[Minor]}}$\xspace}

\newcommand{\perrange}[4]{$\approx #1$~$\pm$~#2\%~$\rightarrow #3$~$\pm$~$#4$\%\xspace}

\title[Galaxy mergers triggering BH growth in \eagle]{Galaxy mergers in \eagle do not induce a significant amount of black hole growth yet do increase the rate of luminous AGN}

\author[S. McAlpine et al.]{Stuart McAlpine$^{1}$\thanks{E-mail:
stuart.mcalpine@helsinki.fi}, Chris M. Harrison$^{2}$, David J. Rosario$^{3}$, David M. Alexander$^{3}$,
\newauthor
Sara L. Ellison$^{4}$, Peter H. Johansson$^{1}$, and David R. Patton$^{5}$
\\
$^{1}$Department of Physics, University of Helsinki, Gustaf H\"{a}llstr\"{o}min katu 2 P.O. Box 64, FI-00014 University of Helsinki, Finland\\
$^{2}$School of Mathematics, Statistics and Physics, Newcastle University, Newcastle upon Tyne, NE1 7RU, UK\\
$^{3}$Centre for Extragalactic
Astronomy, Department of Physics, Durham University, South Road, Durham DH1
3LE, UK\\
$^{4}$Department of Physics \& Astronomy, University of Victoria, Finnerty Road, Victoria, British Columbia, V8P 1A1, Canada\\
$^{5}$Department of Physics and Astronomy, Trent University, 1600 West Bank Drive, Peterborough, ON K9L 0G2, Canada}

\date{Accepted XXX. Received YYY; in original form ZZZ}

\pubyear{2017}

\begin{document}
\label{firstpage}
\pagerange{\pageref{firstpage}--\pageref{lastpage}}
\maketitle

\begin{abstract}

We investigate the connection between galaxy--galaxy mergers and enhanced black hole (BH) growth using the cosmological hydrodynamical \eagle simulation. We do this via three methods of analysis, investigating: the merger fraction of AGN, the AGN fraction of merging systems and the AGN fraction of galaxies with close companions. In each case, we find an increased abundance of AGN within merging systems relative to control samples of inactive or isolated galaxies (by up to a factor of $\approx 3$ depending on the analysis method used), confirming that mergers are enhancing BH accretion rates for at least a subset of the galaxy population. The greatest excess of AGN triggered via a merger are found in lower mass ($M_* \sim 10^{10}$~\Msol) gas rich ($f_{\mathrm{gas}} > 0.2$) central galaxies with lower mass BHs ($M_{\mathrm{BH}} \sim 10^{7}$~\Msol) at lower redshifts ($z<1$). We find no enhancement of AGN triggered via mergers in more massive galaxies ($M_* \gtrsim 10^{11}$~\Msol). The enhancement of AGN is not uniform throughout the phases of a merger, and instead peaks within the early \emph{remnants} of merging systems (typically lagging $\approx 300$~Myr post-coalescence of the two galaxies at $z=0.5$). We argue that neither major ($M_{\mathrm{*,1}} / M_{\mathrm{*,2}} \geq \frac{1}{4}$) nor minor mergers ($\frac{1}{10} < M_{\mathrm{*,1}} / M_{\mathrm{*,2}} < \frac{1}{4}$) are statistically relevant for enhancing BH masses globally. Whilst at all redshifts the galaxies experiencing a merger have accretion rates that are on average 2--3 times that of isolated galaxies, the majority of mass that is accreted onto BHs occurs outside the periods of a merger. We compute that on average no more than 15\% of a BHs final day mass comes from the enhanced accretion rates triggered via a merger.
\end{abstract}


\section{Introduction}
\label{sect:introduction}

\begin{table*}

\caption{An overview of the four samples used  throughout this study, showing their selection criteria, the unique selection criteria of their associated control sample, the mean number of galaxies that meet each selection criteria per simulation output and the results section the sample is used for. For the \squotes{Major mergers} sample, |\ndynmajor| $\leq 1$ refers to galaxies that have completed or will complete a major merger within $\pm 1$ dynamical time, i.e., they are \squotes{in the state of a major merger} (see \cref{sect:eagle}). For the \squotes{Close pairs} sample, $r_{\mathrm{sep[Major]}}$ refers to the 3D distance to the closest major companion. \squotes{Major} in all cases refers to a stellar mass ratio of $M_{\mathrm{*,1}} / M_{\mathrm{*,2}} \geq \frac{1}{4}$, where $M_{\mathrm{*,2}}$ is always set to be the stellar mass of the most massive member of the galaxy pair. Note only galaxies with stellar masses $M_* \geq 10^{10}$ \Msol in the redshift range $0<z<5$ are considered for each sample.}

\begin{tabular}{cccrrrc} \hline

Sample & Selection Criteria & Unique Control Criteria & \multicolumn{3}{c}{$\langle N \rangle$} & Reference\\
& & & $0<z<1$ & $1<z<2$ & $2<z<5$ \\

\hline\hline

AGN luminosity & \Lbol $\geq 10^{43}$ \erg & \Lbol $< 10^{43}$ \erg & 373 & 589 & 269 & \cref{sect:merger_fraction} \\
Eddington rate & \edd $\geq 10^{-2}$ & \edd $< 10^{-2}$ & 263 & 472 & 206 & \cref{sect:merger_fraction} \\
Major mergers & |\ndynmajor| $\leq 1$ & |\ndynmajor| $> 2$ & 410 & 467 & 166 & \cref{sect:agn_fraction} \\
Close pairs & $r_{\mathrm{sep[Major]}} \leq 100$~pkpc & $r_{\mathrm{sep[Major]}} > 200$~pkpc & 424 & 443 & 130 & \cref{sect:close_pairs} \\

\hline
\end{tabular}
\label{table:samples}
\end{table*}

The physical link between actively accreting supermassive black holes (BHs,
referred to as active galactic nuclei, or AGN) and galaxy--galaxy interactions is the subject of a complex and on-going debate, first systematically explored over 30 years ago \citep{Sanders1988}. Theoretically, there are compelling reasons why one would expect such a link to exist.  For example, the strong gravitational torques induced during gas rich major mergers (typically defined as $\leq$ 4:1 stellar mass ratios) can effectively funnel gas toward the nuclei, fuelling bursts of star formation and nuclear activity
\citep[e.g,][]{BarnesAndHernquist1991,MihosAndHernquist1996,Blumenthal2018}.  Additionally, numerical simulations of gas-rich major mergers have shown significant enhancements in star formation \citep[e.g,][]{Johansson2009,Volonteri2015a,Zolotov2015,Pontzen2017} and BH activity for at least one of the systems during the course of the interaction \citep[e.g,][]{DiMatteo2005,Springel2005c}. If the induced growth via the merging process were to contribute a significant fraction to the stellar and BH mass budgets, it could naturally give rise to the empirical scaling relations between the BH mass and various properties of the host galaxy, such as the velocity dispersion and mass of the stellar bulge \citep[e.g,][]{Magorrian1998,McConnellandMa2013}. Alternatively, the induced growth may be entirely non-consequential, with the correlations between BHs and their host galaxies only appearing as result of a random walk \citep[e.g.,][]{Peng2007,Hirschmann2010,Jahnke2011}.

From an empirical point of view, the picture linking galaxy interactions to BH activity is less clear. At higher redshifts ($z \gtrsim 1$), extremely luminous (\Lbol $\geq 10^{46}$ \erg, where \Lbol is the bolometric AGN luminosity) heavily obscured quasars are found to reside almost exclusively in disturbed systems, strongly in line with a merger driven scenario \citep[e.g,][]{Glikman2015,Fan2016}. However, \citet{Schawinski2012} see no such trend for similarly luminous AGN, finding the majority of their host galaxies to be disk dominated, and therefore showing no sign of a recent interaction. Still at high redshift, low and intermediate luminosity AGN (\Lbol $\leq 10^{45}$ \erg) typically exhibit merging fractions very similar to that of the inactive population \citep[e.g,][]{Kocevski2012,Schawinski2011,Rosario2015,Mechtley2016,Marian2019}, suggesting that mergers have little influence towards enhancing BH activity in this regime.

The equivalent empirical studies at lower redshifts ($z \lesssim 1$) are also mixed. \citet{Goulding2018} utilise a novel machine-learning technique applied to over 100,000 spectroscopically confirmed systems in an attempt to automatically identify those with and without merging features. They find galaxies in the current state of a merger are $\approx$ 2--7 times more likely to contain a luminous AGN than their non-interacting counterparts.  This quantitatively agrees with previous studies, who also find a noticeable enhancement in the fraction of AGN that reside in either close pairs or morphologically disturbed hosts above a control sample \citep[e.g,][]{Cotini2013,Ellison2011,Ellison2013,Ellison2015,Koss2010,Satyapal2014,Rosario2015,Koss2018}. Yet, again, many low-redshift studies also fail to find a distinction between the AGN fraction of interacting and non-interacting galaxies \citep[e.g,][]{Cisternas2011,Villforth2014,Hewlett2017,Villforth2017}.  Thus with the potential exception of extremely luminous AGN at high-redshift, it still remains unclear from observations what role galaxy--galaxy mergers have to play in triggering BH activity.  

The discrepancies in the results between observational studies have been attributed to multiple factors. When trying to investigate correlations over a wide dynamic range of AGN luminosities, the small samples sizes of many of these studies can be particularly restrictive.  More fundamentally, dust obscured AGN in merging systems may be missed entirely in surveys that only focus on shorter wavelengths \citep[e.g.,][]{Goulding2009,Weston2017,Koss2018}, indicating that surveys in the infrared and rest frame hard x-rays may be the most effective measure of AGN selection \citep[e.g.,][]{Brandt2015}. Perhaps most crucially, the process of identifying merging systems through morphological disturbances or asymmetry is especially challenging, and often done by eye \citep[however this process is becoming increasingly automated with improving success, e.g.,][]{Pawlik2016,Goulding2018,Bottrell2019}. As the surface brightness of tidal features is intrinsically low, particularly at high redshift and for low mass ratio interactions, many interacting systems may simply be misidentified as non-interacting. Similarly, resolving the final stages of the merger (the coalescence of the two galaxies nuclei), or identifying the signatures of galaxies immediately post-merger, are also extremely challenging, and require sensitive imaging. Finally, when selecting on a variable processes, such as AGN activity, any correlations that exist on average may be washed out entirely \citep{Hickox2014}, suggesting that a selection on both AGN activity and the merging indicators may be required for a fuller understanding \citep[such as was done for][, finding indeed that both mergers have an excess of AGN and AGN hosts are more frequently disturbed]{Ellison2019}.

Hydrodynamical simulations of merging systems have provided compelling theoretical evidence for a link between BH activity and galaxy interactions \citep[e.g.,][]{Dubois2015,Pontzen2017}, yet the global significance of the merging process for boosting BH activity within a full cosmological context remains largely unknown. \citet{Steinborn2018} investigated the role of galaxy mergers as driving mechanisms for BH activity in the high mass regime ($M_* \geq 10^{11}$ \Msol) using the cosmological hydrodynamical {\sc Magneticum Pathfinder} simulations. They argue, that whilst the merger fractions of AGN hosts can be up to three times higher than those of inactive galaxies, the role of mergers in high-mass galaxies are not statistically relevant for BH fuelling.

For this study we utilise \eagle, a cosmological hydrodynamical simulation with more than an order of magnitude higher mass resolution than {\sc Magneticum Pathfinder}, which has proven to reproduce many properties of the observed Universe with high fidelity: such as the colour bimodality of galaxies \citep{Trayford2015}, the evolution of galaxy sizes and star formation rates \citep{Furlong2015,Furlong2017} and the correlation between the star formation rate and BH activity \citep{McAlpine2017,Scholtz2018}. Here we build upon these successes, and investigate the connection between galaxy--galaxy mergers and BH activity.    

The paper is organised as follows. In \cref{sect:eagle} we briefly overview the \eagle simulation, our sample selection and our control pairing criteria. \cref{sect:results} contains our results: investigating the merger fraction of AGN in \cref{sect:merger_fraction}, the AGN fraction of merging systems in \cref{sect:agn_fraction} and the AGN fraction of close pairs in \cref{sect:close_pairs}. We discuss our results, including a comparison to current observational studies, in \cref{sect:discussion}, and finally conclude in \cref{sect:conclusions}.

\section{The \eagle simulation}
\label{sect:eagle}

\eagle \citep[\dquotes{Evolution and Assembly of GaLaxies and their Environment},][]{Schaye2015,Crain2015} \footnote{\url{www.eaglesim.org}}\textsuperscript{,}\footnote{The galaxy and halo catalogues of the simulation suite, as well as the particle data, are publicly available at \url{http://www.eaglesim.org/database.php} \citep{McAlpine2015,EAGLE2017}.} is a suite of cosmological smoothed particle hydrodynamics (SPH) simulations that cover a range of periodic volumes, numerical resolutions and physical models. To incorporate the processes that operate below the simulation resolution a series of \squotes{subgrid} prescriptions are implemented, namely: radiative cooling and photo-ionisation heating \citep{Wiersma2009a}; star formation \citep{Schaye2008}, stellar mass loss \citep{Wiersma2009b} and stellar feedback \citep{DallaVecchia_Schaye2012}; BH growth via accretion and mergers, and BH feedback \citep{Springel2005a,Schaye2015,RosasGuevara2016}. The free parameters of these models are calibrated to reproduce the observed galaxy stellar mass function, galaxy sizes and the BH mass--bulge mass relation at $z \approx 0.1$.  A full description of the simulation and the calibration strategy can be found in \citet{Schaye2015} and \citet{Crain2015}, respectively. 

For this study, we are interested in the influence of galaxy--galaxy mergers as triggering mechanisms for BH activity. Therefore to cover the widest dynamic range of AGN luminosities, Eddington rates and host galaxy diversities, we restrict our study to the largest simulation, denoted Ref-L0100N1504. This simulation is a cubic periodic volume 100 comoving megaparsecs (cMpc) on each side, sampled by $1504^{3}$ dark matter particles of mass $9.7 \times 10^{6}$~\Msol and an equal number of baryonic particles with an initial mass of $1.8 \times 10^{6}$~\Msol.  The subgrid parameters are those of the \eagle reference model, described fully by \citet{Schaye2015}. The simulation adopts a flat $\Lambda$CDM cosmogony with parameters inferred from the analysis of {\it Planck} data \citep{Planck2013}: $\Omega_\Lambda = $\,0.693, $\Omega_{\rm m} = $\,0.307, $\Omega_{\rm b} =$\,0.048, $\sigma_8 =$\,0.8288, $n_{\rm s} = $\,0.9611 and $H_0 = $\,67.77\,km\,s$^{-1}$\,Mpc$^{-1}$. A \citet{Chabrier2003} stellar initial mass function (IMF) is adopted.

The complete state of the simulation is stored at 400 intervals between redshift $z=20$ and $z=0$ in a series of data-lite \squotes{snipshots}. In post-processing, the dark matter structure finding algorithm \dquotes{Friends of Friends} and the substructure finding algorithm \subfind \citep{Springel2001,Dolag2009} were performed on 200 of these outputs to produce a set of halo and galaxy catalogues. The galaxies are then tracked through cosmic time via a merger tree, with the history of each galaxy being considered from the reference frame of their main progenitor, defined as the branch of the galaxy's full merger tree that contains the greatest total mass \citep[see][ for full details]{Qu2017}.

Halo mass, \M{200}, is defined as the total mass enclosed within $r_{\mathrm{200}}$, the radius at which the mean enclosed density is 200 times the critical density of the Universe (i.e., $200 \rho_{\mathrm{crit}})$. Galaxy mass, \M{*}, is defined as the total stellar content bound to a subhalo within a spherical aperture with radius 30~proper kiloparsecs (pkpc), as per \citet{Schaye2015}.

\subsection{The BH subgrid model}
\label{sect:bh_model}

The most influential subgrid models for this study are those that govern the behaviour of BHs, and therefore here we briefly outline their implementation. For a complete description of these models see \citet{Schaye2015} and \citet{RosasGuevara2015}, to see how BHs were considered during the calibration strategy see \cite{Crain2015}.

BHs are initially seeded with a mass of $m_{\mathrm{seed}} = 1.48 \times 10^{5}$~\Msol into dark matter haloes of mass $M_{\mathrm{halo}} = 1.48 \times 10^{10}$~\Msol that do not already contain a BH. The BHs are then free to grow via the Eddington limited accretion of neighbouring gas using a modified Bondi-Hoyle \citep{Bondi1944} formalism that accounts for the angular momentum of the gas \citep{RosasGuevara2015}, i.e.,

\begin{equation}
\dot m_{\mathrm{BH}} = \dot m_{\mathrm{bondi}} \times \mathrm{min}(C^{-1}_{\mathrm{visc}} (c_{\mathrm{s}} / V_{\phi})^{3},1),
\end{equation}

\noindent where $\dot m_{\mathrm{bondi}}$ is the \citet{Bondi1944} rate for spherically symmetric accretion,

\begin{equation}
\label{eq:bondi}
\dot m_{\mathrm{bondi}} = \frac{4 \pi G^{2} m_{\mathrm{BH}}^2 \rho}{(c_s^2 +v^2)^{3/2}}.
\end{equation}

\noindent Here, $m_{\mathrm{BH}}$ is the mass of the BH, $\rho$ is the density of the surrounding gas, $c_s$ is the sound speed of the surrounding gas, $v$ is the relative velocity of the BH and the surrounding gas and $V_{\phi}$ is the rotation speed of the surrounding gas. $C_{\mathrm{visc}}$ is a free parameter related to the viscosity of the (subgrid) accretion disc \citep[see][]{RosasGuevara2015}. BHs also grow via mergers with neighbouring BHs. This occurs instantaneously when two BHs overlap to within each others smoothing kernel (equating to a median  separation of $\approx 1$~pkpc at all redshifts) and their relative velocity to one another is less than the circular velocity at that distance \citep[see][ for a detailed description of this process]{Salcido2016}. The feedback from BHs is implemented using only a single mode, whereby energy is injected thermally and stochastically into the surrounding gas, raising their temperature by a fixed increment.

We note that during the calibration of the subgrid models the observed BH mass--stellar mass relation at $z \approx 0$ was deliberately achieved \citep{Crain2015}. However, the influence of mergers upon BH growth was never considered during this process, and thus is a direct prediction of the simulation. The \eagle simulation under this setup has produced an overall realistic BH population \citep[e.g.,][]{Schaye2015,RosasGuevara2016}, capable of matching many observed relations and behaviours \citep[e.g.,][]{RosasGuevara2016,McAlpine2015,Scholtz2018}.   

\subsection{Galaxy--galaxy mergers}

A galaxy is said to of undergone a merger within the simulation if two independent bound dark matter haloes from a simulation output go on to become a single bound dark matter halo in the next simulation output \citep[bound as defined by the \subfind algorithm, see][ for more details]{Qu2017}. We therefore know the cosmic time of coalescence between two galaxies, denoted $t_{\mathrm{merger}}$, to within the temporal spacing of the simulation outputs (i.e., to within $\approx 50$~Myr), and we assign a random cosmic time between the two outputs for the value of $t_{\mathrm{merger}}$. The mergers between two galaxies are classified by the stellar mass ratio, $M_{\mathrm{*,1}} / M_{\mathrm{*,2}}$, where $M_{\mathrm{*,2}}$ is always set to be the stellar mass of the most massive member of the galaxy pair. A merger is considered to be \squotes{major} if $M_{\mathrm{*,1}} / M_{\mathrm{*,2}} \geq \frac{1}{4}$ and \squotes{minor} if $\frac{1}{10} < M_{\mathrm{*,1}} / M_{\mathrm{*,2}} < \frac{1}{4}$. To account for the stellar stripping that occurs during the later stages of the interaction, the stellar mass ratio is computed when the galaxy in-falling onto the main progenitor had its maximum mass \citep[e.g.,][]{Rodriguez-Gomez2015,Qu2017}. 

Following \cite{McAlpine2018}, we parameterize the \squotes{merging state} of a galaxy by its value of \ndyn, defined as the number of dynamical times to the nearest, i.e., the most proximate in time, merger, i.e.,

\begin{equation}
\label{eq:ndyn}
n_{\mathrm{dyn}} \equiv \frac{t - t_{\mathrm{merger[nearest]}}}{t_{\mathrm{dyn}}}, 
\end{equation}

\noindent where $t$ is the cosmic time at which the galaxy was sampled (i.e., the cosmic time of the simulation output), $t_{\mathrm{merger[nearest]}}$ is the cosmic time of the most proximate in time merger, and $t_{\mathrm{dyn}}$ is the dynamical time at the time $t$, defined as the free-fall time of a dark matter halo, i.e.,

\begin{equation}
t_{\mathrm{dyn}} \equiv \left(\frac{3 \pi}{32 G (200
\rho_{\mathrm{crit}})}\right)^{1/2}.
\end{equation}

\noindent For reference, the dynamical time is $\approx$ 1.6~Gyr at $z=0$, $\approx$ 0.5~Gyr at $z=2$ and $\approx$ 0.2~Gyr at $z=5$. A positive value of \ndyn indicates that the galaxy's most proximate in time merger will complete $n$ dynamical times in the future, whilst a negative value of \ndyn indicates that the galaxy's most proximate in time merger has already completed, and was $n$ dynamical times in the past. If a galaxy has a value \ndynbar $\leq 1$ (i.e., it will complete or has completed a merger within one dynamical time) we define the galaxy to be \squotes{in the state of a merger}. We compute \ndyn separately for the most proximate in time major merger and the most proximate in time minor merger, denoted \ndynmajor and \ndynminor respectively. We chose to operate in a fixed window of dynamical time to define our merging state, over a fixed window of cosmic time, to more fairly compare results from a range of redshifts whilst incorporating the evolving dynamical state of the Universe. We acknowledge that the duration of one dynamical time at low redshift is longer than the timescale that is commonly considered for the direct influence of mergers upon AGN activity \citep[$\approx 0.5$~Gyr, e.g.,][]{Hopkins2008,Johansson2009,Steinborn2018}. However, in this study we do find evidence of AGN enhancement at lower redshifts up to one dynamical time after the merger has completed (see \cref{sect:merger_stages}). Regardless, it should be noted that the choice of dynamical time window has a very limited impact on our overall results (see \cref{sect:choice_of_parameters}).   

\subsection{Sample selection}
\label{sect:sample_selection}

Four mock galaxy samples are constructed for the analysis in \cref{sect:results} (also summarised in \cref{table:samples}):

\begin{enumerate}

    \item\Lbol selected: all galaxies hosting a BH with a bolometric AGN luminosity\footnote{The bolometric AGN luminosity if defined as $L_{\mathrm{bol}} = \epsilon_{r} \dot m_{\mathrm{BH}} c^{2}$, where $c$ is the speed of light, $\dot m_{\mathrm{BH}}$ is the accretion rate of the BH and $\epsilon_{r}$ is the radiative efficiency of the accretion disk, which is assumed to be 0.1 \citep{Shakura1973}.} greater than \Lbol $\geq 10^{43}$ \erg.
    
    \item Eddington rate selected: all galaxies hosting a BH with an Eddington rate\footnote{The Eddington rate is defined as $\lambda_{\mathrm{edd}} = L_{\mathrm{bol}} / L_{\mathrm{edd}}$, where $L_{\mathrm{edd}}$ is the Eddington luminosity.} greater than \edd $\geq 10^{-2}$. 
    
    \item Major mergers: all galaxies currently in the state of a major merger, i.e., those with a value \ndynbarmajor $\leq 1$, where major refers to a stellar mass ratio of $M_{\mathrm{*,1}} / M_{\mathrm{*,2}} \geq \frac{1}{4}$. 
    
    \item Close pairs: all galaxies with a major companion. i.e., those with a companion with a stellar mass ratio of $M_{\mathrm{*,1}} / M_{\mathrm{*,2}} \geq \frac{1}{4}$, within a 3D physical separation of $r_{\mathrm{sep[Major]}} \leq 100$ proper kiloparsecs (pkpc).
\end{enumerate}

\noindent Each sample is designed to investigate how mergers influence BH activity from complementary perspectives, analysed separately in \cref{sect:merger_fraction,sect:agn_fraction,sect:close_pairs}. We limit our selections to the redshift range $0<z<5$. To ensure that minor mergers remain well resolved (for the discussion in \cref{sect:are_mergers_important}), for each sample we only consider galaxies with stellar masses greater than $M_* \geq 10^{10}$ \Msol (i.e., $M_{*,1} \geq 10^{9}$~\Msol). The final samples are constructed by combining the galaxies from each simulation output that lie within the desired redshift range. 

\subsubsection{Constructing a control sample}
\label{sect:control_sample}

In order to establish the influence of galaxy mergers upon enhanced BH activity, for each of the four samples outlined above we require a suitably constructed control. Therefore for each selected galaxy, we assign to it a single randomly selected control galaxy. How one selects the control galaxies is not necessarily straightforward, and must reflect the science question that is being asked. For example, when investigating the influence of mergers for creating active galaxies (i.e., \Lbol $\geq 10^{43}$~\erg or \edd $\geq 10^{-2}$), we wish to contrast the behaviours against a control set of \emph{inactive} galaxies (i.e., \Lbol $< 10^{43}$~\erg or \edd $< 10^{-2}$). In addition, we must ensure that the control galaxies are as similar as possible in their integrated properties to the selected galaxies in order to provide the fairest comparison. Typically, the control galaxies for studies of this nature are only paired on their stellar mass and redshift, to account for the known evolution of the merger fraction with both redshift and mass \citep[e.g.,][]{Rodriguez-Gomez2015,Qu2017}. However, there are many other properties of galaxies that could also influence the growth behaviour of BHs: for example the available gas content, BH mass, or environmental properties: such as the halo mass ($M_{\mathrm{200}}$), or the $N_2$ and $r_2$ parameters \citep[defined for this study as the number of major, $M_{\mathrm{*,1}} / M_{\mathrm{*,2}} \geq \frac{1}{4}$, companions within 2~pMpc, and the distance to the  2$^{\mathrm{nd}}$ closest major companion, respectively, similar to the methods of observational studies, e.g.,][]{Ellison2010,Patton2013,Patton2016}. We note that we would always argue against matching on the SFR, as the SFR of a galaxy can also be enhanced during the merger process \citep[e.g.,][]{RodriguezMontero2019}. 

To ensure that the control galaxies are as similar as possible to the selected galaxies, we opt for the following matching criteria: the control galaxy must be taken from the same simulation output (i.e., it has the same redshift, $z$), have a stellar mass ($M_*$), halo mass ($M_{\mathrm{200}}$), gas mass ($M_{\mathrm{gas}}$), BH mass ($M_{\mathrm{BH}}$) and value of $r_2$ to within 0.05~dex of the selected galaxy, and have a value of $N_2$ to within 5\% of the selected galaxy. On top of this, each sample has an additional unique control criteria condition depending on the science question that is being asked (listed in \cref{table:samples}). If multiple galaxies meet these criteria, one galaxy is selected at random to be the control. However if no suitable control galaxy is found, the matching criteria is progressively loosened by increments of 0.05~dex (5\% for $N_2$), up to a maximum of 0.3~dex (30\% for $N_2$), until a suitable control galaxy is found. If there still remains no suitable control galaxy after this process, then the galaxy is discarded from the sample (resulting in 3--7\% of the sample being discarded depending on redshift). We note that for the close pairs sample we match $r_1$ of the control galaxy to $r_2$ of the selected galaxy \citep[i.e., the distance to the second closest major companion of the selected galaxy must match the distance to the closest major companion of the control galaxy, as per][]{Ellison2010,Patton2013,Patton2016}. 

We acknowledge that these matching criteria are beyond current observational capabilities, but employ them for the analysis in \cref{sect:results} to see what role mergers play in triggering AGN activity using the strictest control sets. We investigate how the choice of matching criteria affects the results in \cref{sect:choosing_control_sample}, and discuss what impact this may have when trying to recover any trends observationally in \cref{sect:observations}.   

The control galaxies matched to the selected galaxies from the four samples are combined to construct four associated control samples, which are designed to trace the underlying merger rate (or AGN fraction) of similar galaxies, whilst remaining as independent as possible from the original sample selection. Any trends that deviate from the trends of the control samples tells us how mergers are influencing BH activity in the simulation.

\subsection{The merger fraction}
\label{sect:merger_fraction_definition}

The merger fraction of AGN is defined as the number of AGN with a value of \ndyn that lie within a chosen window, divided by the total number of AGN, i.e.,

\begin{equation}
\label{eq:merger_fraction}
f_{\mathrm{merger,AGN}} = \frac{N_{\mathrm{AGN}}[a \leq n_{\mathrm{dyn}} \leq b]}{N_{\mathrm{AGN}}},
\end{equation}

\noindent where $a$ and $b$ are the minimum and maximum values of \ndyn that the AGN can have to still be considered in a merging state. Our fiducial values are $a=-1$ and $b=1$, i.e., an AGN is considered to be \squotes{within the state of a merger} if it is within $\pm 1$ dynamical time from coalescence of the two galaxies. The merger fraction of the control sample ($f_{\mathrm{merger,control}}$) is defined in the same manner, now considering what fraction of the associated control galaxies have values of \ndyn between $a$ and $b$ divided by the total number of control galaxies. The excess in the merger fraction is simply the ratio of these two fractions (excess $= f_{\mathrm{merger,AGN}} / f_{\mathrm{merger,control}}$). We note for the figures in \cref{sect:results} we convert $f_{\mathrm{merger,AGN}}$ to a percentage for clarity.

We report errors on the merger fraction as the Poisson error, i.e., the numerator in \cref{eq:merger_fraction} is replaced with the square root. As we quote the merger fraction as integer percentages, any error below 0.5\% is reported as 0\%. 

\section{Results}
\label{sect:results}

\begin{figure*} \includegraphics[width=\textwidth]{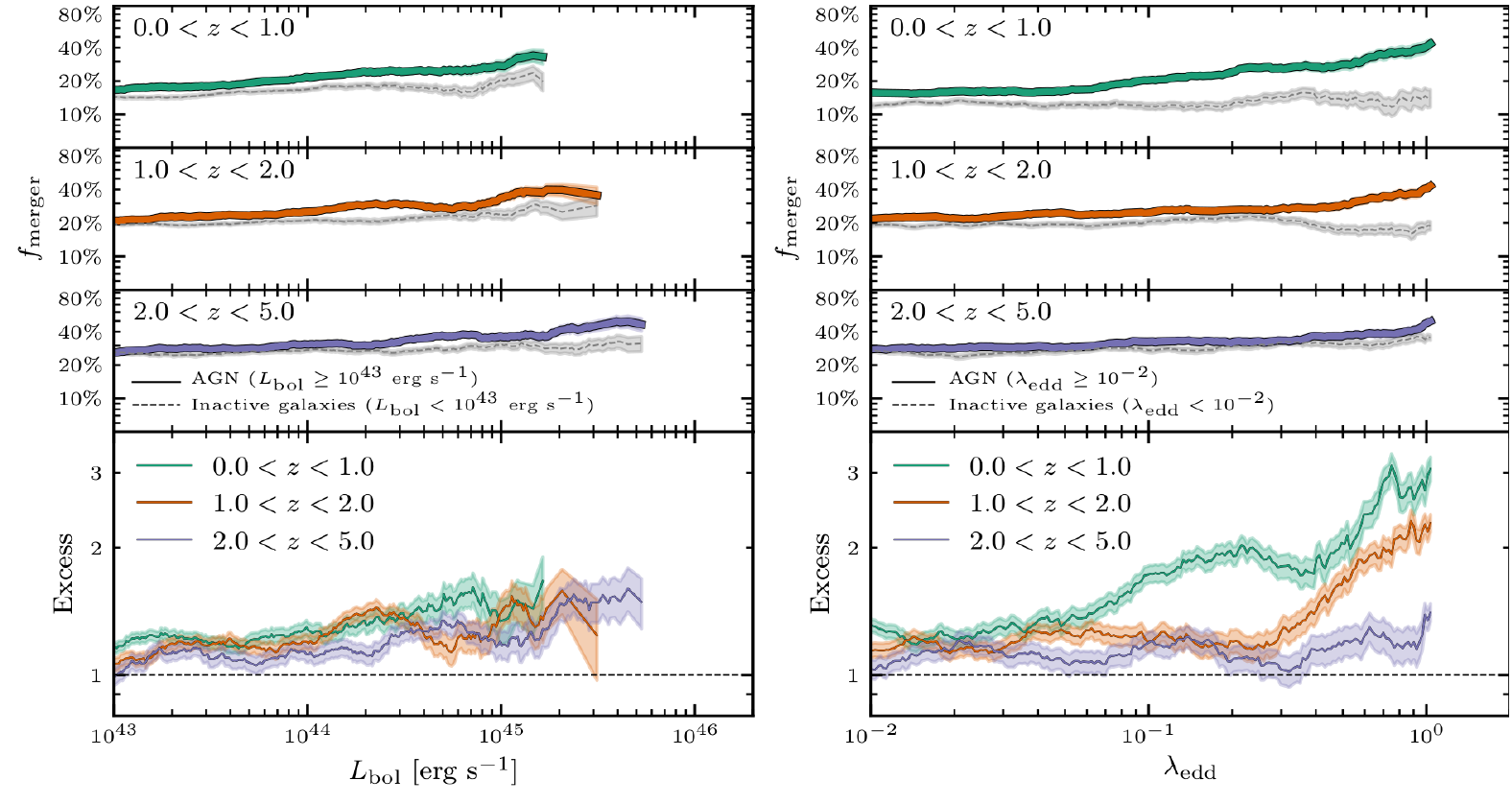}

\caption{The AGN major merger ($M_{\mathrm{*,1}} / M_{\mathrm{*,2}} \geq \frac{1}{4}$) fraction as a function of the bolometric AGN luminosity (left) and the Eddington rate (right). In both panels, the major merger fraction of the associated matched sample of inactive galaxies acts as our control (see \cref{sect:control_sample}). We note that the control galaxies are linked to the galaxies within the active sample using our matching criteria (see \cref{sect:control_sample}), and are presented on the figure using the luminosity or Eddington rate of their associated active galaxy (whilst themselves being inactive galaxies). The shaded regions represent the Poisson uncertainty. For both metrics of BH activity, and for each redshift range, the merger fraction of AGN increases with increasing AGN luminosity or Eddington rate, and is typically higher than the merger fractions of the inactive control samples. The excess between the merger fraction of the AGN and the inactive galaxies is shown in the lower panels: reaching a maximum value of $\approx 1.75$ at high AGN luminosities, and reaching a maximum value of $\approx 3$ at high Eddington rates. This indicates that there is more high-luminosity/Eddington rate AGN in a merging state relative to similar inactive galaxies. This enhancement persists out to the highest redshifts we explore, but is typically more prominent at low redshift (particularly in the case of the Eddington rate). The increased excess when considering the Eddington rate, rather than the AGN luminosity, suggests that it is a clearer indicator of the enhancement of BH activity during mergers.}
\label{fig:major_merger_fraction}
\end{figure*}

\subsection{The enhancement in BH activity due to major mergers}
\label{sect:agn_enhancement}

We begin with an investigation to see if there is a measurable excess in BH activity during the period of a major merger. We do this via three methods, exploring: the merger fraction of AGN in \cref{sect:merger_fraction}, the AGN fraction of merging systems in \cref{sect:agn_fraction}, and the AGN fraction of close pairs in \cref{sect:close_pairs}. Each method tackles the question from a complementary, yet alternative approach, each using a unique galaxy sample and associated control sample, outlined in \cref{table:samples}.

For the analysis below, the samples are split into galaxies at \squotes{low} ($0 < z < 1$), \squotes{intermediate} ($1 < z < 2$) and \squotes{high} redshift ($2 < z < 5$) to avoid misinterpreting any behavioural trend with the underlying evolution of the merger fraction through cosmic time \citep[e.g.,][]{Rodriguez-Gomez2015,Qu2017}. As a reminder, each sample only contains galaxies more massive than $M_* \geq 10^{10}$ \Msol, and here we are only considering the influence of major mergers (i.e., those with stellar mass ratios of $M_{\mathrm{*,1}} / M_{\mathrm{*,2}} \geq \frac{1}{4}$).

We caution that the results from this section should not be directly compared to observational studies in a quantitative sense, as the resulting merger and AGN fractions quoted below are sensitive to our definitions of a \squotes{merging state} and \squotes{active BH} (see \cref{sect:choice_of_parameters}). Furthermore, the galaxy properties chosen to match the selected galaxies to their control galaxies also has an impact on the results (see \cref{sect:choosing_control_sample}), and here were have selected a strict criterion beyond the capabilities of current observational studies. We can however compare the results of this section to observational studies in a qualitative sense, which we discuss in \cref{sect:observations}. In addition, in \cref{sect:obs_comparison_close_pairs} we emulate the observed selection and control pairing criteria of the AGN fractions of close pairs and quantitatively compare the results from the simulation to the observational studies.      

\subsubsection{The merger fraction of AGN}
\label{sect:merger_fraction}

The left panel of \cref{fig:major_merger_fraction} shows the AGN major merger fraction (i.e., the fraction of AGN hosted by galaxies in the state of a major merger) as a function of the bolometric AGN luminosity. Alongside, the merger fraction of the associated control sample of inactive galaxies is also shown. We note that the control galaxies are linked to the galaxies within the active sample using our matching criteria (see \cref{sect:control_sample}), and are presented on the figure using the luminosity or Eddington rate of their associated active galaxy (whilst themselves being inactive galaxies). The merger fraction of the control sample represents the predicted baseline for similar, yet inactive, galaxies, with any deviation from this baseline highlighting the influence of major mergers upon increased BH activity.

The merger fraction of both the AGN and the control galaxies systematically increase with increasing redshift, which is true also for the general population at a fixed mass \citep[$\propto (1+z)^{2.4\text{--}2.8}$, e.g.,][]{Rodriguez-Gomez2015}. Within each redshift range, we find an increasing merger fraction with increasing AGN luminosity: rising from \perrange{17}{0}{33}{5} in the redshift range $0<z<1$, \perrange{21}{1}{40}{5} in the redshift range $1<z<2$, and \perrange{26}{1}{41}{3} in the redshift range $2 < z < 5$ (each reported for the luminosity range $1 \times 10^{43} \leq$ \Lbol $\leq 2 \times 10^{45}$ \erg). The merger fraction of the matched sample of inactive control galaxies similarly increases alongside the AGN sample (due to them being matched on mass, see below). However, due to the shallower gradient in the trends of the control samples, an increasing offset between the two populations emerges. This excess is quantified in the lower panel, showing the ratio of the merger fraction between the active and inactive populations. As the AGN luminosity increases, so too does the excess in the merger fraction, reaching a value of $\approx 1.75$ for the highest luminosities we can explore. This suggests that the brightest AGN, particularly those above \Lbol $\sim 10^{45}$~\erg, reside more often within merging systems over their isolated counterparts. It is also worth noting that even lower luminosity AGN (\Lbol $\sim 10^{43}$~\erg) at lower redshifts ($0<z<1$) exhibit an excess in their merger fraction, which agrees with observations of lower luminosity Seyferts in the local Universe \citep[e.g.,][]{Ellison2011,Ellison2013,Ellison2015}.   

\begin{figure*} \includegraphics[width=\textwidth]{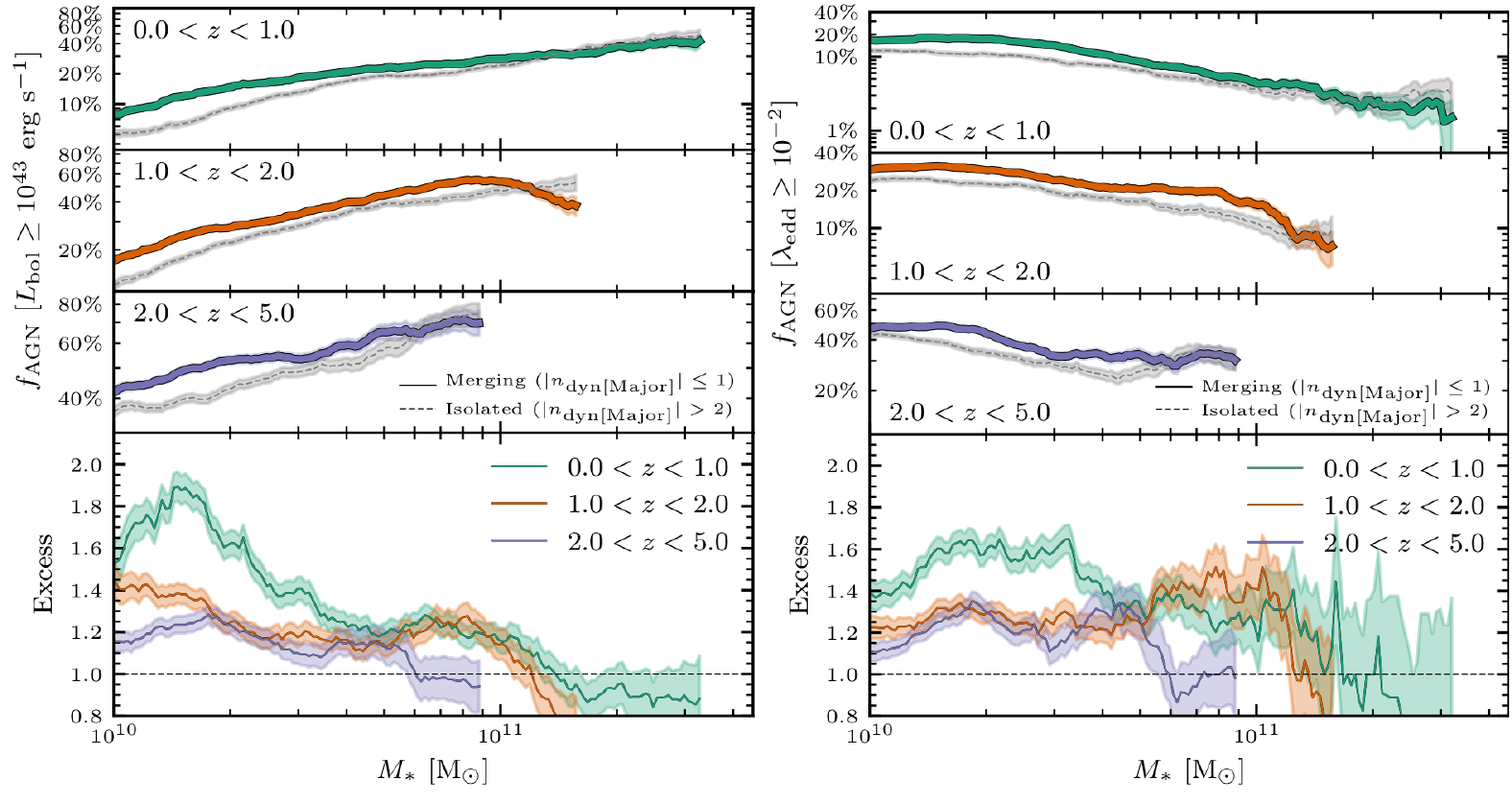}

\caption{The AGN fraction of major mergers ($M_{\mathrm{*,1}} / M_{\mathrm{*,2}} \geq \frac{1}{4}$) as a function of the stellar mass. We classify a galaxy as hosting an AGN if the BH has a bolometric AGN luminosity greater than \Lbol $\geq 10^{43}$ \erg (left panel) or an Eddington rate greater than \edd $\geq 10^{-2}$ (right panel). The AGN fraction of the associated samples of isolated galaxies acts as our control (see \cref{sect:control_sample}). The shaded regions indicate the Poisson uncertainty. As the stellar mass increases, the AGN fraction increases when defined by a cut in AGN luminosity and decreases when defined by a cut in Eddington rate. These trends are due to an increasing BH mass with increasing stellar mass, and are emulated in the trends of the isolated control galaxies. The excess between the AGN fraction of merging systems relative to the isolated control galaxies is shown in the lower panels: at lower redshifts ($z<1$) the excess increases with decreasing stellar mass (up to a value of $\approx 1.8$ at $M_* \approx 10^{10}$~\Msol), at higher redshifts ($z>1$) the excess values maintain an approximately constant value of $\approx 1.1$--1.3 for all stellar masses. However, the spread at higher stellar masses ($M_* \gtrsim 10^{11}$~\Msol) are sufficiently large as to be consistent with no excess (see also \cref{fig:fmerger_fractions_bymass}).} 
\label{fig:agn_fraction}
\end{figure*}

The increasing merger fraction with increasing AGN luminosity seen in the left panel of \cref{fig:major_merger_fraction} is, in part, also driven by mass. This is because whenever we consider the most luminous AGN, we are typically biased towards more massive BHs \citep[which typically reside in more massive galaxies, and the merger fraction of galaxies increases with increasing mass at a fixed redshift, e.g.,][]{Rodriguez-Gomez2015,Qu2017}. Indeed, the BHs in the redshift range $0<z<1$ that occupy the lowest AGN luminosity bin in \cref{fig:major_merger_fraction} have a median mass of $M_{\mathrm{BH}} \approx 3 \times 10^{7}$ \Msol (hosted by galaxies with a median mass of $M_{\mathrm{*}} \approx 3 \times 10^{10}$ \Msol), whereas the BHs in the highest luminosity bin have a median mass of $M_{\mathrm{BH}} \approx 2 \times 10^{8}$ \Msol (hosted by galaxies with a median mass of $M_{\mathrm{*}} \approx 1 \times 10^{11}$ \Msol). This mass differential from the low- to high-luminosity end is why the merger fraction of the (stellar and BH mass matched) control galaxies also increases.

If we consider the mass bias inherent to the AGN luminosities, it is therefore potentially more informative to investigate the major merger fraction of AGN as a function of a BH mass weighted property, such as the Eddington rate. Using this metric, we can more fairly identify the BHs with atypically high (or low) accretion rates, independent of their mass. We investigate the AGN major merger fraction as a function of the Eddington rate in the right panel of \cref{fig:major_merger_fraction}, finding similar overall trends to the bolometric AGN luminosity in the left panel: rising from \perrange{16}{0}{43}{4} in the redshift range $0<z<1$, \perrange{22}{1}{43}{3} in the redshift range $1<z<2$, and \perrange{28}{1}{50}{2} in the redshift range $2<z<5$ (each reported for the Eddington rate range $10^{-2} \leq$ \edd $\leq 1.48$\footnote{The BH accretion rate in the \eagle reference model is capped to the Eddington limit over $h$ (i.e., the maximum allowed value of $\lambda_{\mathrm{edd}} = 1/h = 1.48$).}). The excess in the merger fraction relative to the inactive control galaxies are also similar to the trends found for the bolometric luminosity in the left panel, with the exception of high Eddington rate galaxies at lower redshifts ($z<2$), where the excess now reaches values of $\approx$ 2--3. This increased excess at lower redshifts is in part due to the slight downturn of the merger fraction of the control galaxies beyond \edd $\approx 3 \times 10^{-1}$, created from the increasing dominance of lower mass galaxies over their more massive counterparts with increasing Eddington rate.

We note that the merger fractions in the upper panels of \cref{fig:major_merger_fraction}, and the resulting value of the fractional excess shown in the lower panel of \cref{fig:major_merger_fraction}, are sensitive to our definition of a \squotes{merging state}. In \cref{sect:choice_of_parameters} we explore how the choice of dynamical time window used to define the state of a merger (which for this study was chosen to be $\pm 1$ dynamical time, i.e., $a=-1$ and $b=1$ from \cref{eq:merger_fraction}) influences the resulting excess values, finding that shorter dynamical time windows typically result in larger excess values (by up to a factor of $\approx 2$, but is often much less, see \cref{fig:time_excess_compare}). However, the overall behaviour in the trends (i.e., an increasing excess with increasing AGN luminosity or Eddington rate) is not impacted by the choice of dynamical time window.   

Thus we find that AGN are more commonly found in merging systems over their inactive counterparts. The excess signal is most prominent in two cases: (1) from luminous (\Lbol $\geq 10^{45}$~\erg) massive BHs ($M_{\mathrm{BH}} \sim 10^{8}$~\Msol), where the excess in the merger fraction reaches a factor of $\approx$ 1.75, and (2) from less massive BHs ($M_{\mathrm{BH}} \sim 10^{6}$~\Msol) accreting close to the Eddington limit, where the excess in the merger fraction reaches a value of $\approx$ 1.5--3. Selecting BHs by the Eddington rate appears to provide a fairer view of how mergers influence BH activity, as it can more fairly include the contribution from lower mass BHs/galaxies whose intrinsically low AGN luminosities (but high Eddington rates) are typically lost to the background of regularly accreting more massive BHs/galaxies.

\begin{figure*} \includegraphics[width=\textwidth]{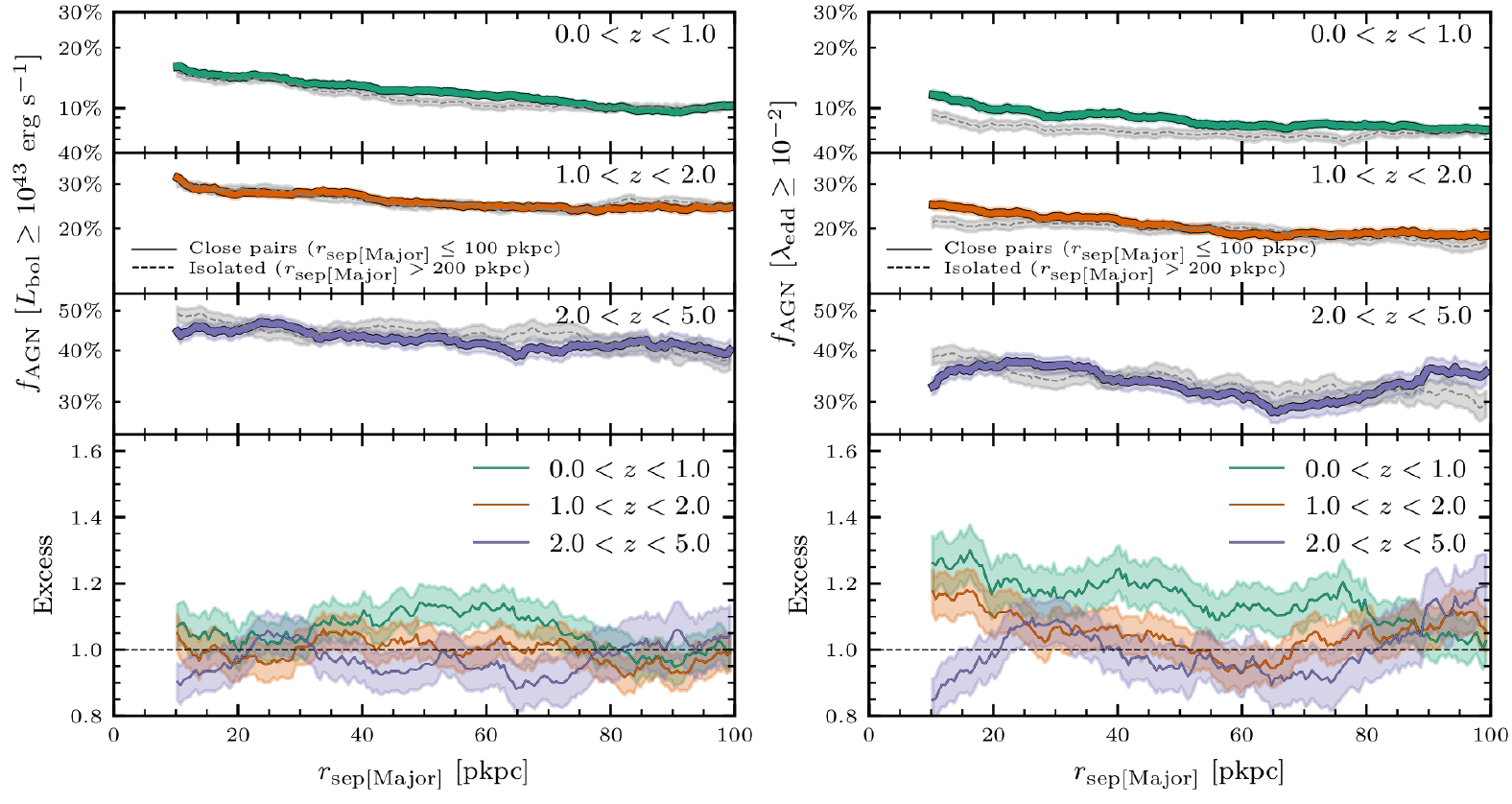}

\caption{The AGN fraction of galaxies with a close major companion ($M_{\mathrm{*,1}} / M_{\mathrm{*,2}} \geq \frac{1}{4}$) as a function of the 3D pair separation. We define a galaxy to be an AGN if it hosts a BH with a bolometric luminosity greater than \Lbol $\geq 10^{43}$ \erg (left panel) or an Eddington rate greater than \edd $\geq 10^{-2}$ (right panel). The AGN fraction of the associated control sample of isolated galaxies is also shown (see \cref{sect:control_sample}). We note that the isolated control galaxies are linked to the galaxies within the close pairs sample using our matching criteria (see \cref{sect:control_sample}), and are presented on the figures using the 3D separation of the close pair galaxy (whilst themselves having no close major companions within 200~pkpc). The shaded regions indicate the Poisson uncertainty. When the AGN fraction is defined by either a cut in the bolometric luminosity or the Eddington rate, there is a weak trend of a rising AGN fraction with decreasing pair separation. The excess in the AGN fraction of close pair galaxies relative to the isolated control galaxies is shown in the lower panels. For the AGN luminosity in the left panel, there is a hint that an excess first appears at separations of $r_{\mathrm{sep[Major]}} \approx 80$~pkpc, and potentially increases towards lower separations up to a maximum value of $\approx 1.1$ (yet the errors mean the excess is often consistent with 1, i.e., no excess). However, when an AGN is defined by a cut in the Eddington rate in the right panel, an increasing excess value with decreasing pair separation is much more prominent: starting at $50 \lesssim r_{\mathrm{sep[Major]}} \lesssim 100$~pkpc and reaching a peak excess of $\approx 1.1$--1.3 at $\approx 10$~pkpc (for $z<2$).}

\label{fig:close_pairs}
\end{figure*}
\subsubsection{The AGN fraction of merging systems}
\label{sect:agn_fraction}

When forming correlations between a stochastic process, such as BH accretion, and a typically stable process, such as the evolution of galaxy wide properties, it has been argued that by initially selecting on the highly-variable process one could inadvertently wash out or dilute any underlying correlations that exist between the two processes \emph{on average} \citep[e.g.,][]{Hickox2014}. For this reason, here we investigate the reverse of our approach in \cref{sect:merger_fraction}, that is, rather than considering the merger fraction of AGN, we now consider the AGN fraction of merging systems. Here, a BH is considered \squotes{active} if it has a bolometric luminosity above \Lbol $\geq 10^{43}$ \erg or an Eddington rate above \edd $>10^{-2}$, although we test how the choice of higher limits affects the results in \cref{sect:choice_of_parameters}. To establish the importance of any discovered trend, we again require a control sample. Therefore for each merging system, we match it with a similar \emph{isolated} control galaxy (see \cref{sect:control_sample}), in order to quantify, at fixed $M_*$ and $z$, how the AGN fractions of merging and isolated galaxies compare.  

The left panel of \cref{fig:agn_fraction} shows, as a function of the stellar mass, the fraction of major mergers  (i.e., |\ndynmajor| $\leq 1$) and isolated systems (i.e., |\ndynmajor| $> 2$) that host a BH with a bolometric AGN luminosity \Lbol $\geq 10^{43}$ \erg. At fixed stellar mass, the AGN fraction of both merging and isolated systems systematically decreases with decreasing redshift \citep[commonly referred to as AGN \squotes{downsizing}, e.g.,][]{Hirschmann2014}. Within each redshift range, the AGN fraction of merging galaxies increases with increasing stellar mass: rising from \perrange{8}{1}{41}{5} in the redshift range $0<z<1$, \perrange{17}{1}{62}{12} in the redshift range $1<z<2$, and \perrange{41}{1}{73}{16} in the redshift range $2<z<5$ (each reported for the stellar mass range $1 \times 10^{10} \leq M_* \leq 1 \times 10^{11}$ \Msol). These upward trends simply reflect the fact that more massive galaxies typically host more massive BHs, and as the mass of the BH increases, a luminosity greater than $10^{43}$ \erg can more easily be achieved. It is for the same reason that an upward trend in the AGN fraction is also emulated by the isolated control galaxies.

The excess between the AGN fraction of merging and isolated systems is shown in the lower panel of \cref{fig:agn_fraction}. For each redshift range, the excess increases with \emph{decreasing} stellar mass, up to a maximum value of $\approx 1.8$ for galaxies with stellar masses $M_* \approx 10^{10}$~\Msol at $z=0$. At higher stellar masses ($M_* \gtrsim 10^{11}$~\Msol), there is little evidence for any excess in the AGN fraction. The potential lack of excess in more massive systems could be caused by our choice of AGN limit, as the most massive BHs residing in the most massive galaxies may simply naturally accrete above \Lbol $\geq 10^{43}$~\erg regardless of the merging state (erasing any excess). This does not appear to be the case, however, as even when the AGN limit is increased, the excess remains primarily in galaxies below $M_* \lesssim 10^{11}$~\Msol (see \cref{fig:agn_excess_compare}).        

If we now consider the fraction of merging and isolated galaxies that host BHs with high Eddington rates (\edd $\geq 10^{-2}$, shown in the right panel of \cref{fig:agn_fraction}), we find a decreasing trend with increasing stellar mass: declining from \perrange{17}{1}{5}{0} in the redshift range $0<z<1$, \perrange{29}{1}{16}{1} in the redshift range $1<z<2$ and \perrange{47}{2}{34}{6} in the redshift range $2<z<5$ (each reported for the stellar mass range $1 \times 10^{10} \leq M_* \leq 1 \times 10^{11}$ \Msol). This trend is formed, again, from that of an increasing BH mass with increasing stellar mass, and whilst high luminosities are common for massive BHs, high Eddington rates become increasingly rare. As with the AGN luminosities, the excess between the merging and isolated systems increases with decreasing stellar mass at lower redshifts ($z<1$), but is approximately constant at all stellar masses at higher redshifts ($z>1$). Unlike in \cref{fig:major_merger_fraction}, where the Eddington rate revealed a larger signal in the excess relative to the AGN luminosity, here both the AGN fraction classified by the AGN luminosity or Eddington rate yield similar values. 

Thus we see further evidence that major mergers trigger an increased amount of BH activity, and, as with \cref{fig:major_merger_fraction}, the excess above the control sample appears to be greatest at lower redshifts ($z<1$). We note that the choice of AGN luminosity or Eddington rate cut used to classify an AGN (which was \Lbol $\geq 10^{43}$ \erg or \edd $>10^{-2}$ in \cref{fig:agn_fraction}) does directly impact the excess values, with higher cuts resulting in a greater excess above the control sample of isolated galaxies (see \cref{fig:agn_excess_compare}). This suggests that the most luminous and highest Eddington rate AGN are more strongly linked with interactions (which was also seen in \cref{fig:major_merger_fraction}).

\subsubsection{The AGN fraction of close pairs}
\label{sect:close_pairs}

Our final method of analysis investigates the AGN fraction of galaxies with a close major companion (i.e., a companion with a stellar mass ratio of $M_{\mathrm{*,1}} / M_{\mathrm{*,2}} \geq \frac{1}{4}$) within  a 3D distance of $r_{\mathrm{sep[Major]}} \leq 100$~pkpc (note that pkpc still refers to proper kiloparsecs and not projected kiloparsecs). For a control, we match each galaxy that has a close major companion to a similar \squotes{isolated} galaxy (i.e., one that does not have a major companion within 200~pkpc, see \cref{sect:control_sample}). We note that the isolated control galaxies are linked to the galaxies within the close pairs sample using our matching criteria (see \cref{sect:control_sample}), and are presented on the figures using the 3D separation of the close pair galaxy (whilst themselves having no close major companions within 200~pkpc). As with the previous section, a galaxy is defined to host an AGN if the BH has a bolometric luminosity in excess of \Lbol $\geq 10^{43}$~\erg or an Eddington rate in excess of \edd $\geq 10^{-2}$ (however we test the effect of different cuts in \cref{sect:choice_of_parameters}). 

The left panel of \cref{fig:close_pairs} investigates the AGN fraction of galaxies with a close major companion as a function of the 3D pair separation, where an AGN is defined by a cut in the bolometric luminosity (i.e., \Lbol $\geq 10^{43}$~\erg). There is a weak trend of an increasing AGN fraction with decreasing pair separation: rising from \perrange{10}{0}{16}{1} in the redshift range $0<z<1$, \perrange{24}{1}{32}{1} in the redshift range $1<z<2$, and \perrange{40}{2}{44}{2} in the redshift range $2<z<5$ (each reported for the 3D separation range $100 \geq r_{\mathrm{sep[Major]}} \geq 10$~pkpc). The galaxies within the control samples exhibit a very similar upward tend with decreasing pair separation, resulting in only a marginal excess between the AGN fraction of the close pair galaxies and the isolated control galaxies (hovering around excess values of $\approx 1.1$ for separations $r_{\mathrm{sep[Major]}} \lesssim 80$~pkpc at $z<1$,  shown in the lower panel). The scenario of the galaxies with the closest companions having the highest AGN fractions would presumably point towards further evidence of a triggering influence of interactions upon enhanced BH activity. However, in this case the dominant reason for an increasing AGN fraction with decreasing pair separation is due to an increasing mean stellar mass and gas fraction with decreasing $r_{\mathrm{sep[Major]}}$, which is why the (stellar and gas mass matched) control galaxies trace the trends so closely. This is caused by the fact that many of the close pair galaxies at larger separations ($r_{\mathrm{sep[Major]}} \gtrsim 50$~pkpc) are gas-poor satellite galaxies hosted within larger haloes ($M_{\mathrm{200}} \sim 10^{13}$~\Msol), whereas at smaller separations ($r_{\mathrm{sep[Major]}} \lesssim 50$~pkpc) the sample begins to become increasingly dominated by interactions between the central galaxies of lower mass haloes ($M_{\mathrm{200}} \sim 10^{12}$~\Msol).  

The right panel of \cref{fig:close_pairs} repeats this analysis for when an AGN is defined by a cut in the Eddington rate (\edd $\geq 10^{-2}$). Again, a weak trend of an increasing AGN fraction with decreasing pair separation is found: rising from \perrange{8}{0}{12}{1} in the redshift range $0<z<1$, \perrange{19}{1}{25}{1} in the redshift range $1<z<2$, and \perrange{34}{2}{33}{2} in the redshift range $2<z<5$ (each reported for the 3D separation range $100 \geq r_{\mathrm{sep[Major]}} \geq 10$~pkpc). The excess in the AGN fraction between the merging and isolated galaxies is much more prominent when the Eddington rate is considered: initially appearing at separations of $50 \lesssim r_{\mathrm{sep[Major]}} \lesssim 100$~pkpc and rising to an excess value of $\approx 1.1$--1.3 at $r_{\mathrm{sep[Major]}} \approx 10$~pkpc (for redshifts $z<2$). At higher redshifts ($z>2$) there is little evidence for any enhancement in the AGN fractions when considering either the bolometric luminosity or the Eddington rate, however we note that the number of galaxies with stellar masses greater than $M_* > 10^{10}$~\Msol that have a major companion at close separations ($r_{\mathrm{sep[Major]}} \lesssim 30$~pkpc) are very limited within the simulation volume at these redshifts.

One could argue that the reduced values of the excess in the AGN fractions seen in \cref{fig:close_pairs} (particularly for the AGN luminosity) are in tension with the results from \cref{fig:major_merger_fraction,fig:agn_fraction}. However, we remind the reader that the close pairs sample is only able to probe galaxies in a pre-merger stage when the two galaxies remain separated, whereas the other three samples additionally include galaxies in a post-merger stage (i.e., any triggered AGN activity post-coalescence is not seen in the close pair analysis, see also \cref{sect:merger_stages}). We additionally note that greater excess values are seen between the AGN fraction of close pair galaxies and their isolated control galaxies if we consider a higher cut in the luminosity or Eddington rate to define an AGN (see \cref{fig:agn_excess_compare_close_pairs}). 

Thus each of the three methods of analyses used in \cref{sect:merger_fraction,sect:agn_fraction,sect:close_pairs} have reported a similar picture, that there exists a measurable excess of AGN activity during the course of a major merger. 

\subsection{The optimal galaxies for enhancing BH activity during a major merger}
\label{sect:by_property}

In the previous section we investigated the merger and AGN fractions for all galaxies more massive than $M_* \geq 10^{10}$~\Msol, discovering a measurable enhancement of BH activity directly connected to the triggering influence of major mergers. To explore this enhancement in more depth, we now test under what conditions the triggering of BH activity during the course of a major merger is optimal. Here we only directly report the results for galaxies within the redshift range $0<z<1$ (where we have the greatest dynamic range of galaxy properties), however we note that the behaviours at higher redshifts are very similar. 

In \cref{fig:agn_fraction} we found an increasing excess in the number of AGN that reside in merging galaxies, relative to the associated control sample of isolated galaxies, with decreasing stellar mass (most strongly at redshifts $0<z<1$). Because of this, we first revisit the results of \cref{fig:major_merger_fraction,fig:close_pairs}, to see if there exists a similar stellar mass dependence upon the excess merger and AGN fractions reported in the lower panels.

\begin{figure} \includegraphics[width=\columnwidth]{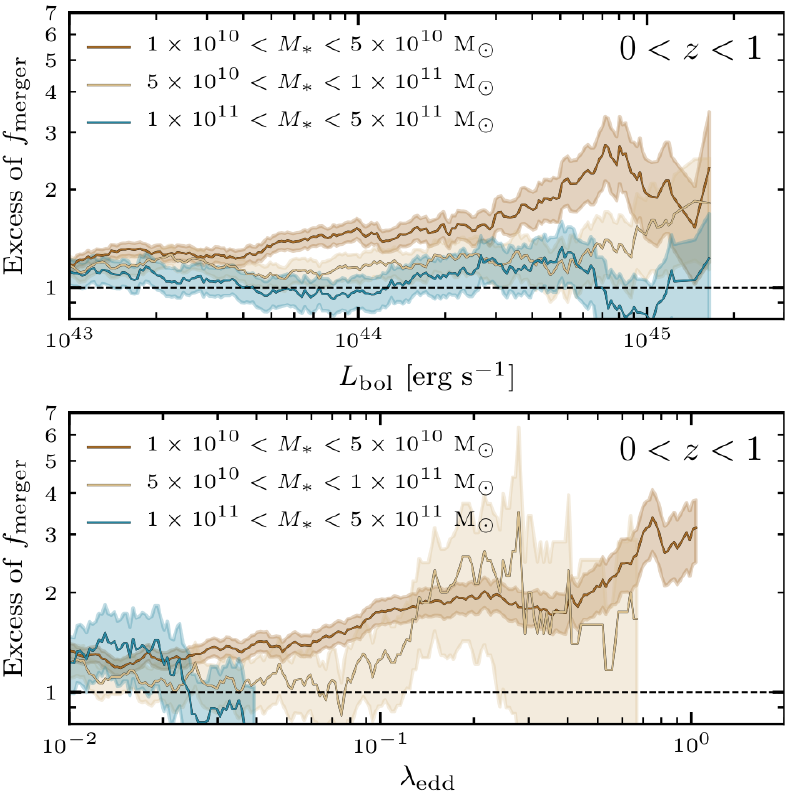} 

\caption{The excess in the major merger fraction from the lower panels of \cref{fig:major_merger_fraction}, with the galaxies in the redshift range $0<z<1$ now subdivided into three stellar mass ranges. Both when investigated as a function of the AGN luminosity (upper panel), or as a function of the Eddington rate (lower panel), the excess in the merger fraction comes almost exclusively from lower mass galaxies ($M_* \lesssim 5 \times 10^{10}$~\Msol). Higher mass galaxies ($M_* \gtrsim 10^{11}$~\Msol) show no excess in their merger fractions between active and inactive galaxies. This suggests that the enhancement of BH activity triggered via a major merger is restricted to less massive systems ($M_* \lesssim 10^{11}$~\Msol).}

\label{fig:fmerger_fractions_bymass}
\end{figure}

The upper panel of \cref{fig:fmerger_fractions_bymass} shows the excess of the merger fraction between active (\Lbol $\geq 10^{43}$~\erg) and inactive BHs (\Lbol $< 10^{43}$~\erg) as a function of the bolometric AGN luminosity (i.e., the lower left panel of \cref{fig:major_merger_fraction}), with the galaxies now subdivided into three stellar mass ranges. It is immediately clear that major mergers do not uniformly enhance BH activity across all of the galaxies within the sample: the BHs hosted by lower mass galaxies ($1 \times 10^{10} <M_* < 5 \times 10^{10}$~\Msol) show the greatest enhancement of BH activity over their inactive counterparts (reaching excess values of $\approx 3$ at $\approx 1 \times 10^{45}$~\erg, over twice the excess that was seen in \cref{fig:major_merger_fraction}), and the most massive galaxies ($M_* \gtrsim 10^{11}$~\Msol) show essentially no enhancement in BH activity over their inactive counterparts. This echoes the results from \cref{fig:agn_fraction}, where the excess in the AGN fraction of merging galaxies over their isolated counterparts was mostly restricted to lower mass systems ($M_* \sim 10^{10}$~\Msol). The lower panel of \cref{fig:fmerger_fractions_bymass} repeats this analysis for the excess of the merger fraction between active (\edd $\geq 10^{-2}$) and inactive BHs (\edd $< 10^{-2}$) as a function of the Eddington rate (i.e., the lower right panel of \cref{fig:major_merger_fraction}), finding that the largest excess values, and the BHs with the highest Eddington rates, are again almost exclusively found in lower mass systems ($M_* \lesssim 5 \times 10^{10}$~\Msol).

\begin{figure} \includegraphics[width=\columnwidth]{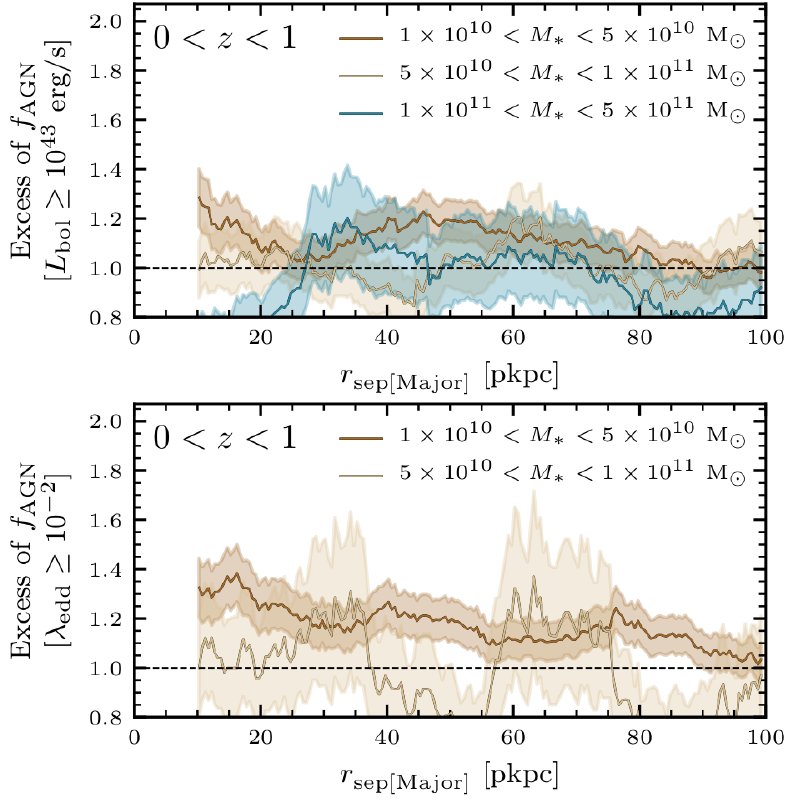} 

\caption{The excess in the AGN fraction from the lower panels of \cref{fig:close_pairs}, with the galaxies in the redshift range $0<z<1$ now subdivided into three stellar mass ranges. When an AGN is classified by either a cut in the bolometric AGN luminosity (\Lbol $\geq 10^{43}$~\erg, upper panel), or by a cut in the Eddington rate (\edd $\geq 10^{-2}$, lower panel), the largest enhancement in the AGN fraction is found in lower mass systems ($M_* \lesssim 5 \times 10^{10}$~\Msol). Higher mass galaxies ($M_* \gtrsim 1 \times 10^{11}$~\Msol) with close major companions show no sign of any excess in their AGN fractions over their isolated counterparts. This suggests that the enhancement of BH activity triggered via a major merger is restricted to less massive systems ($M_* \lesssim 10^{11}$~\Msol).}

\label{fig:closepairs_fractions_bymass}
\end{figure}

\begin{figure} \includegraphics[width=\columnwidth]{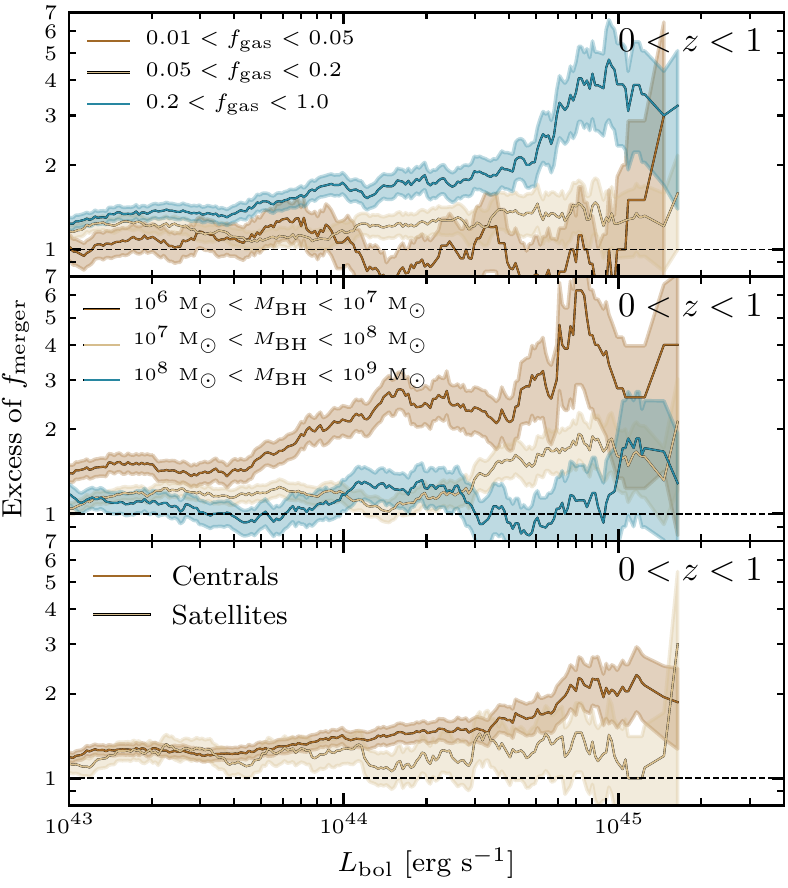} 

\caption{The excess in the major merger fraction from the lower left panel of \cref{fig:major_merger_fraction}, with the galaxies in the redshift range $0<z<1$ now subdivided into ranges of the total gas fraction ($f_{\mathrm{gas}} \equiv \frac{M_{\mathrm{gas}}}{M_{\mathrm{gas}} + M_*}$, upper panel), the BH mass (middle panel) and distinguishing between central and satellite galaxies (lower panel). We find the largest excess between the merger fraction of AGN and inactive galaxies comes from the galaxies that are the most gas rich ($f_{\mathrm{gas}} > 0.2$), those that host lower mass BHs ($M_{\mathrm{BH}} \sim 10^{6}$~\Msol), and from those that are central galaxies.}

\label{fig:fractions_byproperty}
\end{figure}

In a similar manner, \cref{fig:closepairs_fractions_bymass} returns to the analysis of \cref{fig:close_pairs}, investigating the excess between the AGN fractions of galaxies with close major companions and isolated galaxies as a function of the pair separation, now in three bins of stellar mass (we note there are too few galaxies above $M_* \geq 10^{11}$~\Msol within the close pair sample to retrieve meaningful statistics when the Eddington rate is considered). The upper panel of \cref{fig:closepairs_fractions_bymass} classifies an AGN by a cut in the bolometric AGN luminosity (\Lbol $\geq 10^{43}$~\erg, i.e., the lower left panel of \cref{fig:close_pairs}) and the lower panel of \cref{fig:closepairs_fractions_bymass} classifies an AGN by a cut in the Eddington rate (\edd $\geq 10^{-2}$, i.e., the lower right panel of \cref{fig:close_pairs}). Whilst not as elevated as the excess values in \cref{fig:fmerger_fractions_bymass}, we similarly find that less massive systems with close major companions are the ones with the largest excess in their AGN fractions over their isolated counterparts, and, again, the most massive galaxies ($M_* \gtrsim 5 \times 10^{10}$~\Msol) show little evidence for any enhancement in their AGN fractions over their isolated counterparts. 

In addition to the stellar mass, investigating further properties of galaxies may continue to refine what are the optimal conditions for triggering BH activity during a major merger. \cref{fig:fractions_byproperty} again shows the excess of the major merger fraction from the lower left panel of \cref{fig:major_merger_fraction}, with the galaxies now subdivided into ranges of the total gas fraction ($f_{\mathrm{gas}} \equiv \frac{M_{\mathrm{gas}}}{M_{\mathrm{gas}} + M_*}$, upper panel), the BH mass (middle panel) and distinguishing between central and satellite galaxies (lower panel). Intuitively, the merging galaxies with the highest gas fractions ($f_{\mathrm{gas}} > 0.2$) show the greatest excess values in their merger fractions above their inactive counterparts. In addition, we find that the galaxies hosting less massive BHs ($M_{\mathrm{BH}} < 10^{7}$~\Msol) display the greatest excess values, in line with the picture that less massive galaxies are those with the highest excess values (see \cref{fig:fmerger_fractions_bymass,fig:closepairs_fractions_bymass}). Finally, central galaxies appear responsible for much of the excess, as opposed to gas-poor satellite galaxies, particularly at higher AGN luminosities.

It is not entirely clear why BH activity triggered via a merger should be restricted to galaxies of lower masses ($M_* \lesssim 10^{11}$~\Msol). The simplest explanation is that for a galaxy to sustain an AGN for a period of time it requires an adequate supply of fuel (i.e., a high gas fraction), most commonly present in lower-mass galaxies. Additionally, the higher-mass BHs occupying higher-mass galaxies could more rapidly extinguish continued accretion over their lower-mass counterparts via efficient AGN feedback (given the larger accretion rates achieved by more massive BHs under the same surrounding gas conditions, see \cref{eq:bondi}).      

Thus the results from \cref{sect:agn_enhancement,sect:by_property} have shown that major mergers do trigger an increased amount of AGN activity within the \eagle universe, and that it is most measurable at the highest AGN luminosities (\Lbol $\sim 10^{45}$~\erg) and Eddington rates (\edd $\approx 1$), within lower mass \emph{central} galaxies ($M_* \lesssim 10^{11}$~\Msol) with higher gas fractions ($f_{\mathrm{gas}} > 0.2$) that host lower mass BHs ($M_{\mathrm{BH}} \sim 10^{6}$~\Msol) at lower redshifts ($z<1$). 

\subsection{The enhancement of BH activity during different stages of a major merger}
\label{sect:merger_stages}

\begin{figure} \includegraphics[width=\columnwidth]{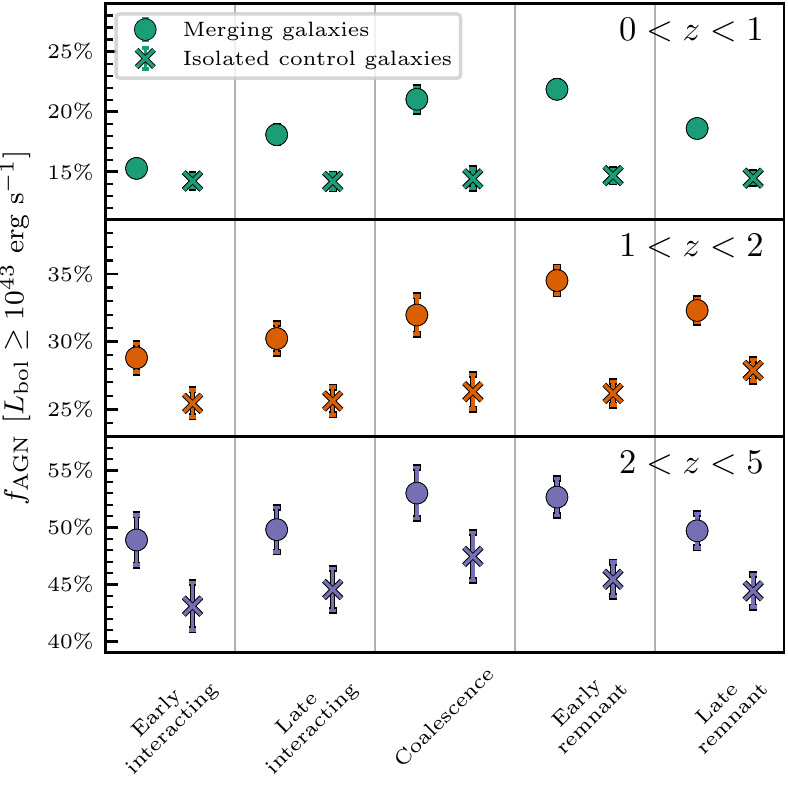} 

\caption{The AGN fraction of galaxies at five predefined stages of a major merger. Each stage is defined using a fixed window of \ndynmajor, i.e., a fixed window of the number of dynamical times to the coalescence of the two galaxies (see \cref{sect:merger_stages}). At each redshift the AGN fraction rises from the early interacting stage to the early remnant phase, and declines towards the late remnant phase. To compare, the AGN fraction of the matched sample of isolated control galaxies is also shown, revealing that there is typically an increased amount of BH activity at most stages of a major merger relative to their isolated counterparts. The greatest excess of AGN activity during a merger is during the early remnant phase (i.e., after the two galaxies have already coalesced, see also \cref{fig:excess_vs_ndyn}). The error bars indicate the Poisson uncertainty.}

\label{fig:excess_vs_ndyn_broad}
\end{figure}

\begin{figure} \includegraphics[width=\columnwidth]{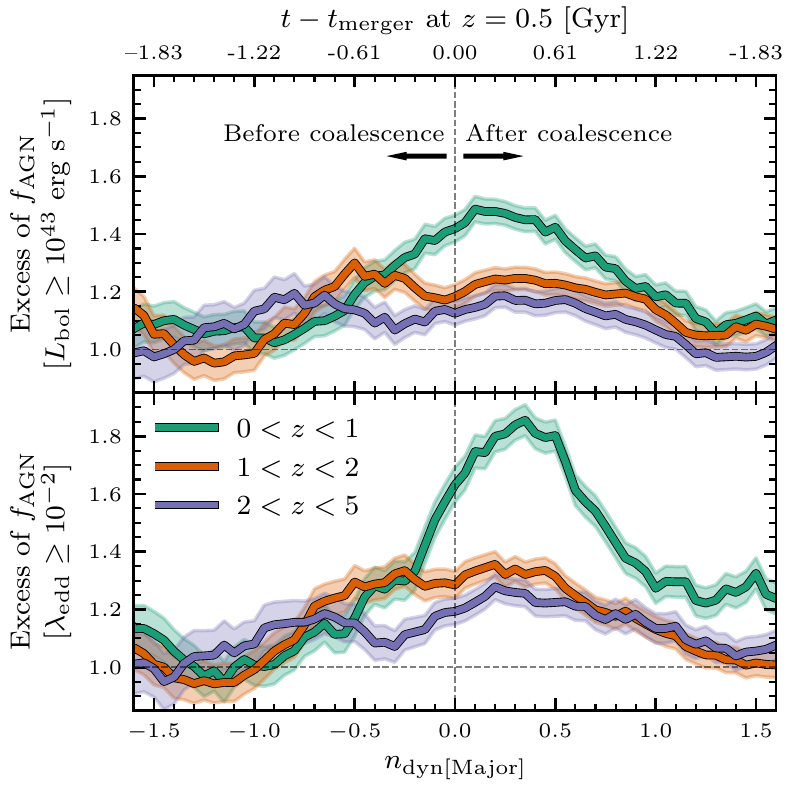} 

\caption{The excess in the AGN fraction (defined by a cut in the luminosity in the upper panel and by a cut in Eddington rate in the lower panel) at each stage of a major merger (parameterized by the number of dynamical times to the coalescence of the two galaxies, i.e., \ndynmajor) relative to the AGN fraction of the associated control sample of isolated galaxies. Here we are showing when during a major merger BH activity is most \emph{enhanced}. Positive values of \ndynmajor indicate the system is post-coalescence, negative values of \ndynmajor indicate the system is pre-coalescence and \ndynmajor values of $\approx 0$ indicate the system is in the final stages of coalescence. At higher redshifts ($z>1$), an excess in the AGN fraction first appears $\approx 1$ dynamical time before the coalescence of the two galaxies, maintains a value of 1.2--1.3 until $\approx 1$ dynamical time after the coalescence of the two galaxies, and declines towards higher values of \ndynmajor. At these redshifts ($z>1$) the total excess in the AGN fraction originates from galaxies both before and after coalescence, with an approximately equal weighting (50/50). At lower redshifts ($z<1$), the majority of the excess in the AGN fraction originates from galaxies post-coalescence, and the distribution is peaked around \ndynmajor $\approx 0.25$, which corresponds to $\approx 300$~Myr of cosmic time at $z=0.5$. This indicates that a significant fraction ($\approx$ 65--75\%) of BH activity that is triggered via a merger occurs within the remnants of merging systems at $z<1$.}

\label{fig:excess_vs_ndyn}
\end{figure}

For the analysis in \cref{sect:agn_enhancement} we only considered our fiducial definition of a merging system: a galaxy is in the state of a merger if it has completed or will complete a major merger within $\pm 1$ dynamical time (see \cref{sect:merger_fraction_definition}). However such a broad time window will shield the relative importance of each merger stage for enhancing BH activity (e.g., the interacting, coalescence and remnant phases). To explore this, we now investigate the AGN fraction of galaxies at various stages of a major merger (parameterized by the number of dynamical times to the coalescence of the two galaxies, i.e., \ndynmajor), to see when, if at all, an optimal stage for triggering BH activity exists. Here we use the galaxies from the \squotes{Major mergers} sample (see \cref{table:samples}).

\cref{fig:excess_vs_ndyn_broad} shows the AGN fraction of galaxies at five predefined stages of a major merger, starting from the initial interaction through to the final remnant. We categorise each merger stage using a fixed window of \ndynmajor, i.e., a fixed window of the number of dynamical times to the coalescence of the two galaxies: \squotes{early interacting} $\equiv -1.0 < n_{\mathrm{dyn[Major]}} < -0.5$, \squotes{late interacting} $\equiv -0.5 < n_{\mathrm{dyn[Major]}} < -0.1$, \squotes{coalescence} $\equiv -0.1 < n_{\mathrm{dyn[Major]}} < 0.1$, \squotes{early remnant} $\equiv 0.1 < n_{\mathrm{dyn[Major]}} < 0.5$ and \squotes{late remnant} $\equiv 0.5 < n_{\mathrm{dyn[Major]}} < 1.0$\footnote{For galaxies at $z=0.5$ these time dynamical time windows correspond to cosmic time windows of: \squotes{early interacting} $\equiv -1.22 < t - t_{\mathrm{merger}} < -0.61$~Gyr, \squotes{late interacting} $\equiv -0.61 < t - t_{\mathrm{merger}} < -0.12$~Gyr, \squotes{coalescence} $\equiv -0.12 < t - t_{\mathrm{merger}} < 0.12$~Gyr, \squotes{early remnant} $\equiv 0.12 < t - t_{\mathrm{merger}} < 0.61$~Gyr and \squotes{late remnant} $\equiv 0.61 < t - t_{\mathrm{merger}} < 1.22$~Gyr. At higher and lower redshifts the dynamical time windows will correspond to shorter and longer cosmic time windows, respectively (see \cref{eq:ndyn}).}. That is, we redefine the values of $a$ and $b$ in \cref{eq:merger_fraction} to these new limits. The AGN fraction of the matched isolated control galaxies associated with the merging galaxies at each stage is also shown. We find, for each redshift range, that the AGN fraction is not constant throughout the merger process, and instead slowly rises and declines throughout the course of the interaction, peaking during the early remnant phase. This tells us that the greatest abundance of AGN during a major merger are found soon after the two galaxies have already coalesced. If we then compare the AGN fractions of the merging systems to the isolated control galaxies, we also find that the greatest \emph{enhancement} of AGN is during the early remnant stage (most notably in the lower two redshift ranges, $z<2$).

Taking this investigation further, \cref{fig:excess_vs_ndyn} shows, now purely as a function of the number of dynamical times to coalescence (i.e., no predefined phases), the \emph{excess} in the AGN fraction of galaxies at a particular stage in a major merger relative to the AGN fraction of their associated isolated control galaxies. As a reminder: negative values of \ndynmajor indicate the nearest major merger is in the future and the system is still in an interacting/pre-coalescence phase, positive values of \ndynmajor indicate the nearest major merger occurred in the past and the system is in a remnant/post-coalescence phase, and values very close to zero indicate the system is in the final stages of coalescence. In the upper panel a galaxy is defined to be active if it has a bolometric AGN luminosity greater than \Lbol $\geq 10^{43}$~\erg (the same as \cref{fig:excess_vs_ndyn_broad}), and in the lower panel a galaxy is defined to be active if it has an Eddington rate greater than \edd $\geq 10^{-2}$.

For the higher two redshift ranges ($z>1$), and for both definitions of an active BH (\Lbol $\geq 10^{43}$~\erg or \edd $\geq 10^{-2}$), an excess in the AGN fraction first appears $\approx 1$~dynamical time ($\approx 1.2$~Gyr at $z=0.5$) before the coalescence of the two galaxies, oscillates steadily around excess values of 1.2--1.3 until 1 dynamical time after the coalescence of the two galaxies, and then continues to decline towards higher values of \ndynmajor. If we integrate under the curve between the limits $-1 <$ \ndynmajor $< 1$ (i.e., our definition of a merging state) we find a very similar total excess both before and after the coalescence of the two galaxies. This means that $\approx 50$\% of the excess values at $z>1$ reported in \cref{fig:major_merger_fraction,fig:agn_fraction} originate from the remnants of merging galaxies. The behaviour changes somewhat at lower redshifts ($z<1$), now with the majority of enhanced BH activity triggered via the merging process occurring after the coalescence of the two galaxies (65\% and 75\% in the upper and lower panels respectively, again in the limits $-1 <$ \ndynmajor $< 1$). This means that at lower redshifts a significant majority of the excess values reported in \cref{fig:major_merger_fraction,fig:agn_fraction} originate from the remnants of merging galaxies. In addition, the distribution at lower redshifts is distinctly peaked around a value of \ndynmajor $\approx 0.25$, corresponding to $\approx 300$~Myr of cosmic time at $z=0.5$, suggesting there is typically a significant delay between the coalescence of the two galaxy nuclei and triggered BH activity at $z<1$.   

\section{Discussion}
\label{sect:discussion}

\subsection{The effect of the model}
\label{sect:model}

When analysing the results from cosmological hydrodynamical simulations, such as \eagle, it is always important to consider how the adopted subgrid models may influence the interpretation of the results. For this study, the most relevant subgrid models are those that govern the behaviour of BHs, which we briefly described in \cref{sect:bh_model}, and are fully described in \citet{Schaye2015}. 

The accretion rate of BHs in the simulation is directly proportional to the density of the surrounding gas, and the square of the mass of the BH ($\dot m_{\mathrm{BH}} \propto m_{\mathrm{BH}}^{2} \rho$, see \cref{eq:bondi}). Thus a high accretion rate can be created as the density of the surrounding gas increases, for example as it is funnelled inward or compressed during the course of a merger, or simply by having a massive BH. Each of these two routes can readily produce visibly \squotes{active} BHs, and both contribute to the upward trends found in \cref{fig:major_merger_fraction,fig:agn_fraction}. These effects are not necessarily contentious, as we would expect better fuelled and larger BHs to be increasingly capable of producing more luminous AGN. However, given that both a jump in the surrounding gas density during a merger versus there simply being an already massive BH are degenerate to the eventual accretion rate, it is not always straightforward to decouple the dominant contributor to any increased AGN activity.

The dependence between the accretion rate of the BH and the square of the BH mass will, at least in part, be responsible for the increased excess of AGN activity seen \emph{after} the coalescence of the two galaxies has completed (see \cref{fig:excess_vs_ndyn_broad,fig:excess_vs_ndyn}). This results from the fact that as the two BHs eventually coalesce (following the coalescence of the two galaxies), the sudden jump in BH mass will result in an even greater jump in the accretion rate (assuming the same conditions of the surrounding gas), increasing the likelihood for a \squotes{visible} AGN in the merger remnant. Additionally, the characteristic timescale between the two galaxies coalescing and the eventual coalescence of the two BHs is dependent on the BH merging criteria adopted by the simulation. For \eagle, the coalescence of two BHs is not a resolved process, and we therefore implement broad conditions for this process to occur: the two BHs must be within each others smoothing kernel and their relative velocity to one another must be less than the circular velocity at that distance. It is likely that these conditions merge the BHs earlier than they should \citep[e.g.,][]{Rantala2017}, which would potentially result in a rightward shifting of the excess peaks in \cref{fig:excess_vs_ndyn} (i.e., the peak of AGN activity would lag further behind the coalescence of the two galaxies). However, we do not anticipate any of the overall behaviour or trends of this study would be affected by this, with the majority of the triggered AGN activity still occurring post-coalescence (of the galaxy nuclei). 

\subsection{Comparing to observations}
\label{sect:observations}

Both at lower and higher redshifts, as of yet there remains no unanimous consensus as to the importance of galaxy--galaxy mergers for triggering BH activity from observational data. A possible exception is the most luminous (\Lbol $\gtrsim 10^{46}$~\erg), typically heavily obscured quasars, which are found to reside almost exclusively in disturbed systems, suggesting a merger driven scenario at least in this regime \citep[e.g.,][]{Glikman2015,Fan2016}. However one should be careful on how to interpret systems with such strong selection biases. Yet even amongst the uncertainty that has arisen between the empirical results, it is still informative to compare the results of the simulation to the observations where possible, along with making predictions for future observations.

For this study we have deliberately chosen to avoid a quantitative comparison with observations when a \squotes{merging state} has to be defined, such as for the results in \cref{sect:merger_fraction,sect:agn_fraction}. In the simulation we have the advantage of knowing when two galaxies will, or have, coalesced, which we parameterized by the number of dynamical times to that event. However observational works must ascertain the merging state of a galaxy from only an instantaneous (often pre-coalescence) snapshot. Thus a truly fair comparison would require us to apply observational techniques to synthetic images to estimate a galaxy's current merging state \citep[similar to][, for example]{Lahen2018,Bottrell2019,Snyder2019}, but this is beyond the scope of this study. We can, however, qualitatively compare our results to the observational studies.  

The trend of an increasing merger fraction with increasing AGN luminosity, similar to the trends found in \cref{fig:major_merger_fraction}, has been discovered empirically \citep[e.g.,][]{Ellison2019}. More broadly, the observed fraction of merging galaxies that host an AGN, or the fraction of AGN found to reside in merging systems, are often reported to be higher than the samples of inactive or isolated control galaxy counterparts \citep[e.g.,][]{Koss2010,Ellison2011,Rosario2015,Goulding2018}. These observations therefore agree with the results presented in \cref{fig:major_merger_fraction,fig:agn_fraction}, and suggests, both in the observations and the simulation, that mergers are directly responsible for triggering an increased amount of BH activity for at least a subset of the galaxy population. However, these observational results, and therefore our own, then disagree with the empirical studies that find no discernible enhancement in AGN activity around the time of a merger \citep[e.g.,][]{Kocevski2012,Schawinski2015,Villforth2017,Marian2019}. When it comes to the observed AGN fraction of galaxies with close companions, an increasing excess of AGN with decreasing pair separation has been found  \citep[e.g.,][]{Ellison2011,Silverman2011}, further reinforcing the mergers triggering BH activity scenario, and again agreeing with the results from this study (see \cref{fig:close_pairs} and also \cref{sect:obs_comparison_close_pairs}). Thus qualitatively the results presented by this study are in good agreement with many current observational works that have investigated the merger--AGN connection.

For future observations, we predict that the strongest observable signal connecting enhanced BH activity to galaxy--galaxy mergers will come more from high Eddington rate sources, as opposed to high luminosity sources \citep[such as was done in][]{Marian2019}. In addition, we predict that the excess in the merger and AGN fractions will be greatest at lower redshifts (i.e., $z<1$), and the galaxies exhibiting the most optimal conditions for triggering an AGN via a merger are those with lower masses ($M_* \sim 10^{10}$~\Msol), higher gas fractions ($f_{\mathrm{gas}} \geq 0.2$) and lower mass BHs ($M_{\mathrm{BH}} \sim 10^{7}$~\Msol, see \cref{fig:agn_fraction,fig:fmerger_fractions_bymass,fig:closepairs_fractions_bymass,fig:fractions_byproperty}). For the three methods of analysis used in \cref{sect:results}, we consistently found no enhancement of BH activity during the period of a merger in the most massive galaxies ($M_* \gtrsim 10^{11}$~\Msol), relative to their inactive or isolated counterparts \citep[where some observational studies have reported their \emph{strongest} signals of AGN enhancement, e.g.,][]{Goulding2018}.  

A key finding of this study was discovering that 50--75\% of enhanced BH activity triggered by major mergers comes after the two galaxies have already coalesced (see \cref{fig:excess_vs_ndyn_broad,fig:excess_vs_ndyn}). Indeed, a much weaker excess was found in the AGN fraction over their isolated counterparts if we restricted our sample to just the galaxies currently in their interacting stages (see \cref{fig:close_pairs}), which is the stage where the majority of observational samples will be capturing galaxies in the state of a merger. It is therefore crucial that observational studies are able to robustly identify post-merger remnants, so as to not mistakenly classify these AGN as being hosted by isolated systems. Encouragingly, studies have shown that post-merger features could have observability time-scales of $\approx 0.2$--0.4~Gyr \citep[e.g.,][]{Lotz2010,Ji2014}, which would mean that the peak excess of BH activity that arises $\approx 300$~Myr after the coalescence of the two galaxies found in \cref{fig:excess_vs_ndyn} at $z<1$ could be captured, and therefore would be correctly attributed to a post-merger system. Indeed, some observational results have reported that the largest excess of AGN activity has been found in post-merger systems \citep[e.g.,][]{Ellison2013,Koss2018}, in agreement with the findings of this study. 

As a final note, we investigated in \cref{sect:choosing_control_sample} how the choice of matching criteria to select the control galaxies could affect the results of studies of this nature. For this study we matched the control galaxies using the redshift, stellar mass, halo mass, BH mass, gas mass and the environment (through the $r_2$ and $N_2$ parameters), to ensure that the control galaxies were as similar as possible to the selected galaxies (see \cref{sect:control_sample}). However, these criteria cannot be trivially adopted for observations, with the majority opting to match on just the redshift and the stellar mass. Generally, we found that when fewer parameters are considered in the matching criteria, the excess values of both the merger fraction of AGN and the AGN fraction of merging systems (i.e., the lower left panels of \cref{fig:major_merger_fraction,fig:agn_fraction}) are typically \emph{higher} (see \cref{fig:excess_compare}). This could imply that observational studies that only match their control galaxies on the stellar mass and redshift are slightly overestimating their values of the excess fractions. However, the behaviours of the loosest control matching criteria are consistent with the strictest control matching criteria, and the excess values are never more than 50\% different (and often much less, see \cref{fig:excess_compare}). Larger differences are seen in the excess fractions when the Eddington rate is considered, varying by up to a factor of two in the excess values between the loosest and strictest matching criteria (see \cref{fig:excess_compare}). This is because the control population becomes biased relative to the galaxies within the selected samples when matched on fewer parameters, caused by the fact that high Eddington rate AGN BHs are typically undermassive for galaxies of their stellar mass. It therefore appears that the excess values when considering the Eddington rate could be rather overestimated when the control galaxies are not matched on the BH mass. Finally, if the environment is not considered (through the $r_2$ and $N_2$ parameters) when investigating the AGN fraction of close pairs (i.e., \cref{fig:close_pairs}), the AGN fraction of the control galaxies can be overestimated, resulting from gas-poor satellite galaxies of massive haloes getting mistakenly assigned to gas-rich central galaxies of the same mass. Yet overall, whilst it is recommended to match on as many parameters as possible, the \emph{behaviours} recovered for each analysis method are largely unaffected by the choice of matching criteria, and the excess values are often well within 50\% of one another between the loosest and strictest matching criteria.

\subsubsection{Directly comparing to observations of the AGN fraction of close pairs}
\label{sect:obs_comparison_close_pairs}

\begin{figure} \includegraphics[width=\columnwidth]{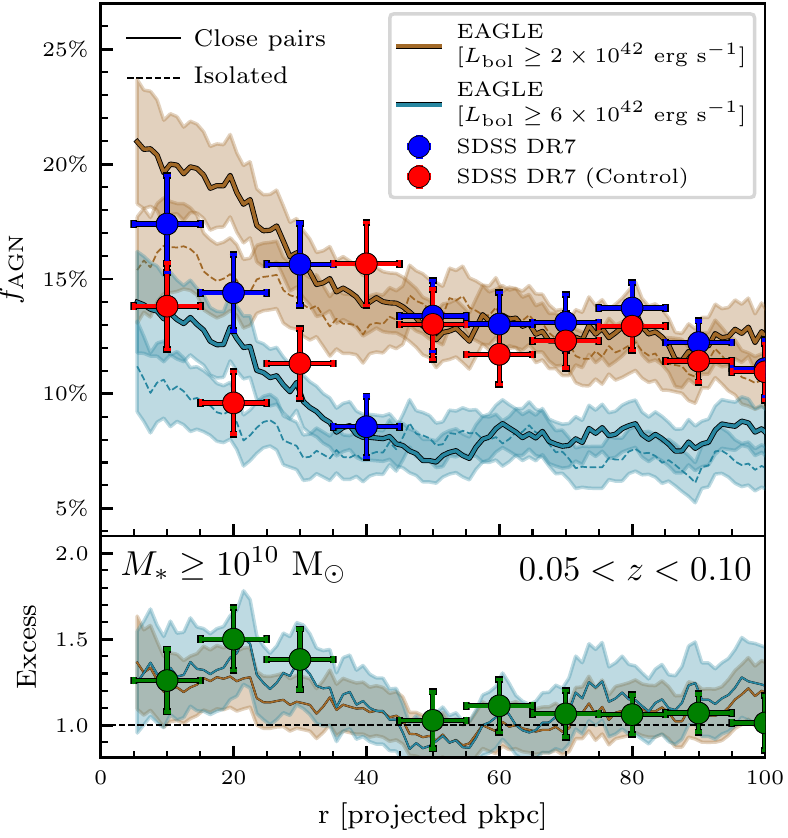}

\caption{The AGN fraction of galaxies with close major companions ($M_{\mathrm{*,1}} / M_{\mathrm{*,2}} \geq \frac{1}{4}$) as a function of the projected pair separation compared to observations taken from the \textit{Sloan Digital Sky Survey} Data Release 7 (SDSS DR7). For both the simulation and the observations, the AGN fractions of the associated isolated control galaxies are also shown (see \cref{sect:control_sample}). For \eagle, a BH is classified as \squotes{active} if it has a bolometric AGN luminosity greater than either of the two quoted cuts. For the observed sample, a BH is classified as \squotes{active} based on the cut of \citet{Kauffmann2003}, with a S/N $>3$ required for all the requisite diagnostic emission lines. Only galaxies more massive than $M_* \geq 10^{10}$~\Msol in the redshift range $0.05 < z < 0.10$ are considered for each sample. There is an increasing AGN fraction with decreasing pair separation below $r \lesssim 40$~projected kpc, and the excess between the AGN fraction of the close pair galaxies and their associated isolated control galaxies (shown in the lower panel) also only exists below these separations (up to a excess value of $\approx 1.5$, albeit with large errors).}

\label{fig:ellison_compare}
\end{figure}

To conclude this section, we examine how the results from the \eagle simulation quantitatively compare to the observations of galaxies in the local Universe with close major companions taken from the \textit{Sloan Digital Sky Survey} Data Release 7 (SDSS DR7). The observed galaxies are classified as hosting an AGN based on the cut of \citet{Kauffmann2003}, with a S/N $>3$ required for all the requisite diagnostic emission lines. The SDSS sample consists of 7,216 galaxies above a stellar mass of $M_* \geq 10^{10}$~\Msol in the redshift range $0.05 <z < 0.10$ that have a close major ($M_{\mathrm{*,1}} / M_{\mathrm{*,2}} \geq \frac{1}{4}$) companion within a separation of 100~projected kpc and a relative velocity to within $\Delta v \leq 300$~km/s. For this analysis, the control galaxies from both the observations and the simulation are matched on redshift, stellar mass and the environment (through the $r_2$ and $N_2$ parameters), using the same method and tolerance levels as outlined in \cref{sect:control_sample}. We note, that when applying the same selection to the simulation, the stellar mass distributions between the observed and simulated samples are not the same, with the simulated galaxy sample containing a greater proportion of lower mass galaxies ($M \sim 10^{10}$~\Msol). To ensure that this does not have an impact on the results, we have rerun the analysis whereby we mass match the galaxies from the simulation to the observations in each bin of projected separation, indeed finding no significant change in the result.    

The comparison is shown in \cref{fig:ellison_compare}, showing the AGN fraction of galaxies with close major companions as a function of the projected separation in the upper panel, and the excess between the AGN fraction of the close pair galaxies and the AGN fraction of their associated isolated control galaxies in the lower panel (analogous to \cref{fig:close_pairs}). As we cannot classify if a galaxy hosts an AGN in the same manner as the observations, we perform the analysis with two cuts in the bolometric AGN luminosity to define an \squotes{active} BH: \Lbol $\geq 2 \times 10^{42}$~\erg and \Lbol $\geq 6 \times 10^{42}$~\erg. These cuts have been chosen to match the normalisation of the observed AGN fractions, and also to demonstrate how sensitive the AGN fractions are to this choice. The behaviour of the simulation for both cuts of AGN luminosity are very similar, showing a rising AGN fraction with decreasing pair separation, only differing from one another by their overall normalisation. This rising trend is also apparent in the observed sample, however the rise in the AGN fraction at smaller separations ($r \lesssim 20$~projected kpc) is potentially less steep in the observations when compared to the simulation (yet remain consistent to within the errors). Focusing now on the excess in the lower panel, we find very similar behaviours for both cuts of AGN luminosity from the simulation and also from the galaxies within the observed sample. At larger separations ($r \gtrsim 40$~projected kpc) there is no notable excess in the AGN fraction relative to their isolated control galaxies, but at smaller separations ($r \lesssim 40$~projected kpc) a trend of a rising excess with decreasing pair separation begins to appear, reaching excess values of around $\approx 1.5$ (albeit with large errors).  

Therefore the results from the simulation and the observations are encouragingly alike, both showing a quantitatively similar degree of evidence for an increased amount of AGN activity for galaxies with close major companions, consistent with the overall results of this study.    

\begin{figure} \includegraphics[width=\columnwidth]{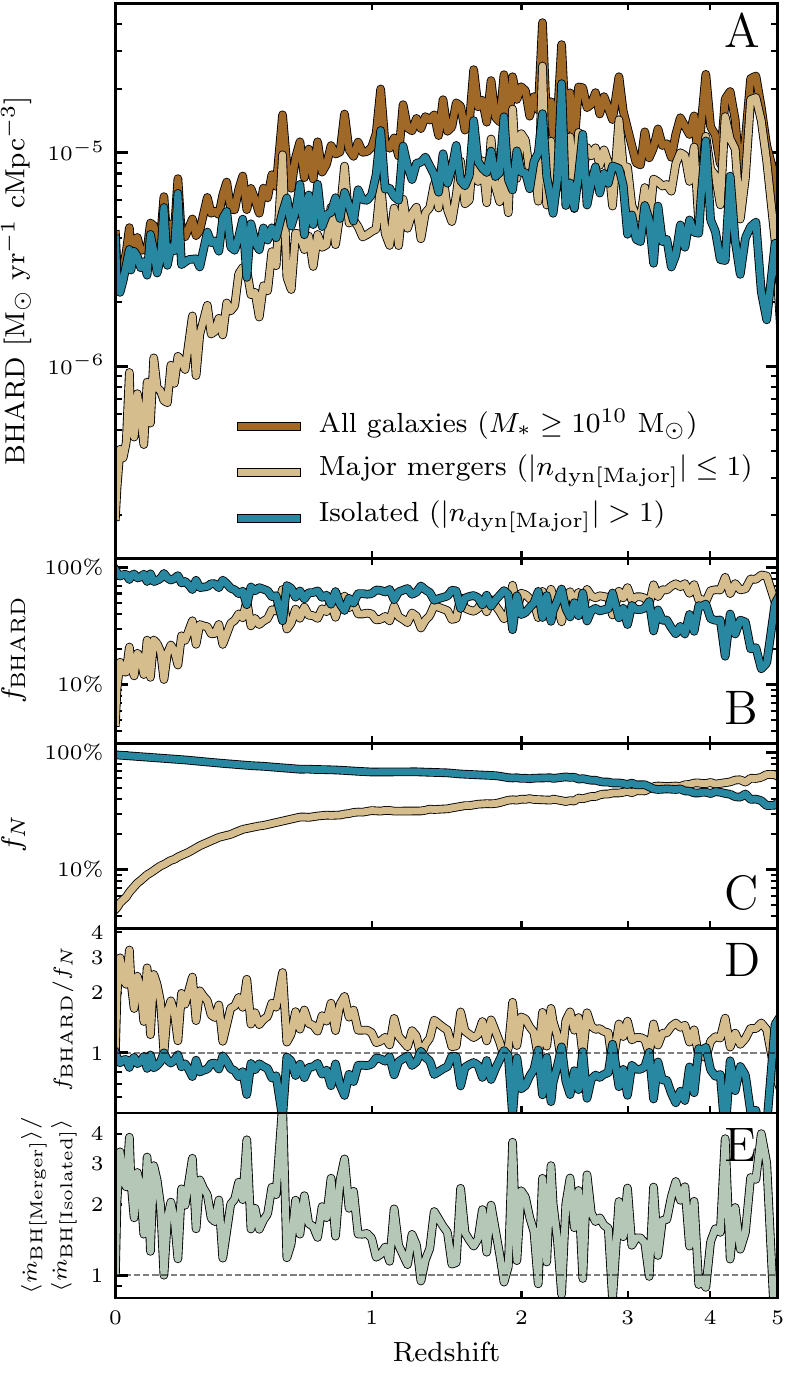}

\caption{\textit{Panel A}: the BH accretion rate density (BHARD) from all galaxies more massive than $M_* \geq 10^{10}$~\Msol, galaxies currently in the state of a major merger (i.e., \ndynmajorbar $\leq 1$) and galaxies not currently in the state of a major merger (i.e., \ndynmajorbar $> 1$). \textit{Panel B}: the fraction of the total BHARD coming from merging and non-merging systems. \textit{Panel C}: the fraction of these galaxies that are in a major merger and not in a major merger. \textit{Panel D}: the ratio between panels \textit{B} and \textit{C}, i.e., the contribution to the BHARD from merging and non-merging systems weighted by their number, or equivalently, the mean accretion rate of merging and isolated galaxies versus the mean accretion rate of all galaxies. \textit{Panel E}: the ratio between the mean accretion rate of merging galaxies and the mean accretion rate of isolated galaxies. At higher redshifts ($z \gtrsim 2$) the majority of the BHARD come from merging galaxies, however the majority of galaxies at this time are in mergers. At lower redshifts ($z \lesssim 2$) both the galaxy population and the BHARD are dominated by isolated systems. At all redshifts, merging galaxies have accretion rates $\approx$2 times greater than isolated galaxies.}

\label{fig:cbhard}
\end{figure}

\begin{figure} \includegraphics[width=\columnwidth]{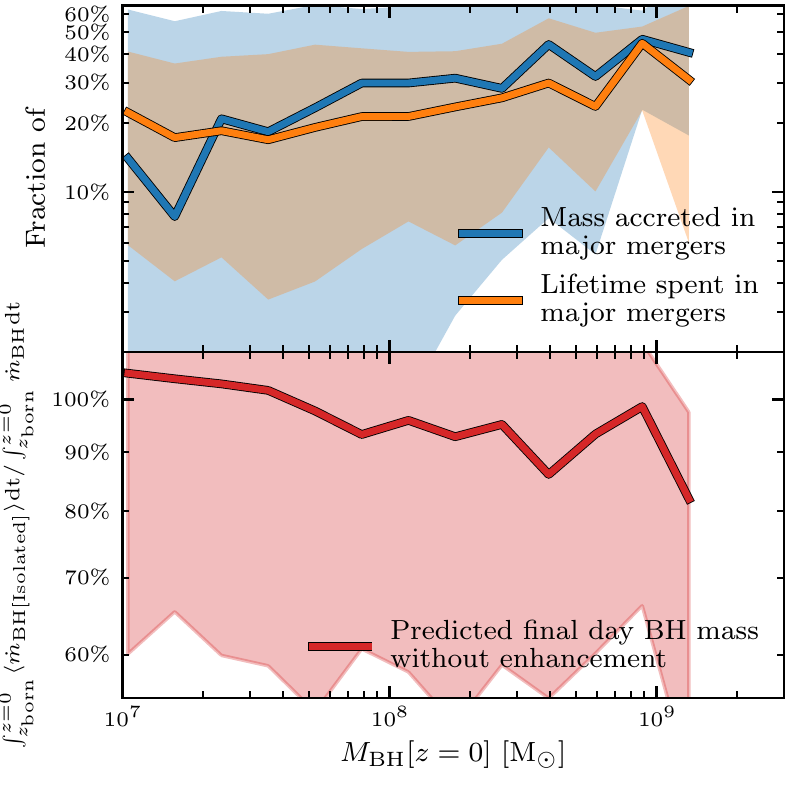}

\caption{\textit{Upper panel}: the fraction of the total \emph{accreted mass} (i.e., $\int_{z_{\mathrm{born}}}^{z=0} \dot m_{\mathrm{BH}}$ dt) that was accreted during the period(s) of a major merger (i.e., the fraction of mass that was accreted within $\pm 1$ dynamical time of the coalescence of the two galaxies) as a function of the present day BH mass. To compare, the fraction of the BHs lifetime that was spent in a major merging system is also shown. On average, BHs accumulate an increasing amount of their accreted mass with increasing present day BH mass during the period(s) of a major merger ($\approx 40$\% at $M_{\mathrm{BH[z=0]}} = 10^{9}$~\Msol), but also spend an increasing fraction of their lifetimes within a merging system with increasing present day BH mass. During the period(s) of a major merger, the average BH never accumulates more than 50\% of their accreted mass. \textit{Lower panel}: the predicted final day BH mass if BHs did not experience any enhancement in their growth triggered via a major merger, relative to the true final day BH mass. The predicted mass is obtained by multiplying the average BH accretion rate during times of isolation by the total lifetime of the BH. On average, we predict BHs would still be $\gtrsim 85$\% of their true mass if mergers did not enhance BH activity.}

\label{fig:fraction_in_mergers}
\end{figure}

\subsection{Is the enhancement of BH activity during major mergers important for BH growth?}
\label{sect:are_mergers_important}

In this study we have investigated the relationship between galaxy--galaxy mergers and enhanced BH activity within a cosmological context. We have found that there exists a measurable excess in the fraction of highly accreting BHs that reside in major mergers relative to those that reside in isolated systems, through both the merger fraction of AGN and the AGN fraction of merging systems. However, it remains difficult to gauge from the values of the fractional \squotes{excess} alone how important major mergers are for producing luminous or high Eddington rate AGN, and if the enhanced BH growth resulting from this process is statistically meaningful. Or, more fundamentally, would the BH population today look the same in a Universe free from any major interactions \citep[which can now be investigated for the evolution of individual systems, e.g.,][, but not for global populations]{Pontzen2017}. We note that when we refer to an enhancement of BH growth, here we are referring to the increased accretion onto BHs directly triggered by the merger process, and not the growth resulting from the coalescence of two BHs.  

Panel \textit{A} of \cref{fig:cbhard} shows the cosmic black hole accretion rate density (BHARD) from all galaxies more massive than $M_* \geq 10^{10}$~\Msol in the \eagle simulation, showing also the contribution from the subset of these galaxies currently in the state of a major merger (i.e., \ndynmajorbar $\leq 1$) and those not currently in the state of a major merger (i.e., \ndynmajorbar $> 1$). At the highest redshifts ($z \geq 3$), the galaxies currently experiencing a major merger contribute the greatest amount to the total BHARD ($\gtrsim 70$\%, panel \textit{B}). However we note that at these redshifts the majority of galaxies above $M_* \geq 10^{10}$~\Msol are in a merging state (panel \textit{C}). During intermediate redshifts ($z \approx 2$) both merging and isolated systems contribute a similar amount to the total BHARD, even although the majority of systems by this time are not experiencing a major merger. As we evolve towards the present day ($z=0$), isolated systems have come to dominate both the galaxy population by number ($\approx 97$\%, panel \textit{C}) and the contribution to the total BHARD ($\approx 90$\%, panel \textit{B}).

The galaxies currently experiencing a major merger always contribute more to the total BHARD \emph{relative} to their abundance, i.e., the ratio between $f_{\mathrm{BHARD}}$ and $f_N$ is always $>1$ (panel \textit{D}). In other words, the average accretion rate of merging galaxies is always higher than the average accretion rate of all galaxies (i.e., $\langle \dot m_{\mathrm{BH[Mergers]}} \rangle / \langle \dot m_{\mathrm{BH[All]}} \rangle > 1$), growing from a factor of $\approx 1.2$ at higher redshifts ($z>1$) up to a factor of $\approx 3$ at $z=0$ (panel \textit{D}). Relative to the accretion rate of isolated galaxies however, merging galaxies are \emph{always} accreting on average at a 2--3 times higher rate (panel \textit{E}). This could suggest that a significantly increased amount of BH growth can be attributed to the triggering influence of major mergers, particularly at higher redshifts ($z \gtrsim 2$) where merging systems are the most abundant. 

From \cref{fig:cbhard} we discovered that merging galaxies at all redshifts accrete at an average rate that is 2--3 times higher than that of isolated galaxies. However, to establish the cumulative impact of this enhancement upon the resulting BH growth we must look at the BH accretion rate \emph{histories} of galaxies. In \cref{fig:fraction_in_mergers} we show the fraction of the total \emph{accreted mass} (i.e., $\int_{z_{\mathrm{born}}}^{z=0} \dot m_{\mathrm{BH}}$ dt) that was accreted during the period(s) of a major merger (i.e., the fraction of mass that was accreted within $\pm 1$ dynamical time of the coalescence of the two galaxies) as a function of the present day BH mass. Although there is an extremely large scatter, the average BH with a present day mass of $M_{\mathrm{BH}} = 10^{7}$~\Msol accumulated $\approx 10$\% of their accreted mass during the period of a major merger, and this number rises to $\approx 40$\% for BHs with a present day mass of $M_{\mathrm{BH}} = 10^{9}$~\Msol. To put this in perspective, we additionally show what fraction of the BHs lifetime was spent in a \squotes{merging state}, revealing a similar rising trend, but a slightly lower normalisation to the mass fractions ($\approx 5$--10\%). This reveals yet more evidence that BHs are accreting proportionally more during their time within a major merger over when they are isolated. 

Exploring the BH accretion rate histories further, we can crudely attempt to estimate what mass a BH would have been if it had never experienced the enhanced accretion rates during a major merger. To do this we evaluate a new present day BH mass by simply multiplying the average accretion rate a BH has over its lifetime during isolation (i.e., at all times it is not in a major merger) by the total lifetime of the BH. We then compare this \squotes{non-enhanced} BH mass ($\int_{z_{\mathrm{born}}}^{z=0} \langle \dot m_{\mathrm{BH[Isolated]}} \rangle$ dt) to the true BH mass ($\int_{z_{\mathrm{born}}}^{z=0} \dot m_{\mathrm{BH}}$ dt) in the lower panel of \cref{fig:fraction_in_mergers}. We find that, whilst the scatter is again large, if a BH was to grow at their mean isolated accretion rate it would typically result in a BH that grows to over $\gtrsim 85$\% of the true mass. Or, said in reverse, on average the cumulative result of the enhanced accretion rates triggered via major mergers are responsible for no more than 15\% of the final BH masses at $z=0$ \citep[strongly in line with the conclusions reported by][, who also find the majority of BH growth from the cosmological hydrodynamical {\sc Horizon-AGN} simulation occurs outwith mergers]{Martin2018}.      

Thus it remains difficult to definitively state the \squotes{importance} of major mergers for enhancing BH growth, yet we would argue that overall they are not statistically relevant fuelling mechanisms for BHs. Major mergers do increase the average accretion rates of BHs at all redshifts, by a factor of 2--3 over their isolated counterparts. However this enhancement is either not great enough, or BHs simply do not experience enough cumulative time in a merging state to feel this enhancement in their final BH mass, with the majority of accreted BH mass being accumulated in an isolated state. It is plausible that mergers do become increasingly important for triggering BH activity with decreasing redshift, as we have seen multiple times throughout this study. However, by these times merging systems are now so rare that their (albeit enhanced) contribution is still not highly significant, and isolated galaxies remain the dominant source of BH accretion at lower redshifts. The conclusion of mergers never being statistically relevant fuelling mechanisms for BHs is consistent with the results from the {\sc Magneticum Pathfinder} simulation, who performed a similar analysis to this study in the high-mass regime \citep[$M_* \geq 10^{11}$~\Msol,][]{Steinborn2018}. However we emphasise that if we were to of restricted our study to just the high-mass regime ($M_* \geq 10^{11}$~\Msol) as they did, we would \emph{not} of found the same result (see \cref{fig:fmerger_fractions_bymass}). 

Even if major mergers are not important for BH growth as a whole, they could still remain important drivers for rare, or unique, events. For example, it is plausible that extremely luminous quasars (\Lbol $\gtrsim 10^{46}$~\erg) cannot be sustained via secular processes, and could therefore require a triggering interaction to occur (we have seen evidence in this study that the most luminous AGN are those most commonly found in merging systems, e.g., \cref{fig:major_merger_fraction,fig:agn_excess_compare}). In addition, in \citet{McAlpine2017} we found that the initiation of the \squotes{rapid growth phase} of BHs was commonly found to occur in close proximity to a merger, and, using a sample of control galaxies, found that the importance of mergers for triggering the rapid growth phase increased with decreasing redshift. This directly agrees with the results presented by this study. The BHs experiencing their rapid growth phase are essentially unhindered in their growth, and as such grow close to the Eddington limit. Therefore it is plausible that the strongest signal in the merger fraction excess using the Eddington rates seen in \cref{fig:major_merger_fraction,fig:agn_fraction,fig:excess_vs_ndyn} is largley from the BHs currently experiencing their rapid growth phase.    

\subsubsection{Are minor mergers important?}
\label{sect:minor_mergers}

\begin{figure} \includegraphics[width=\columnwidth]{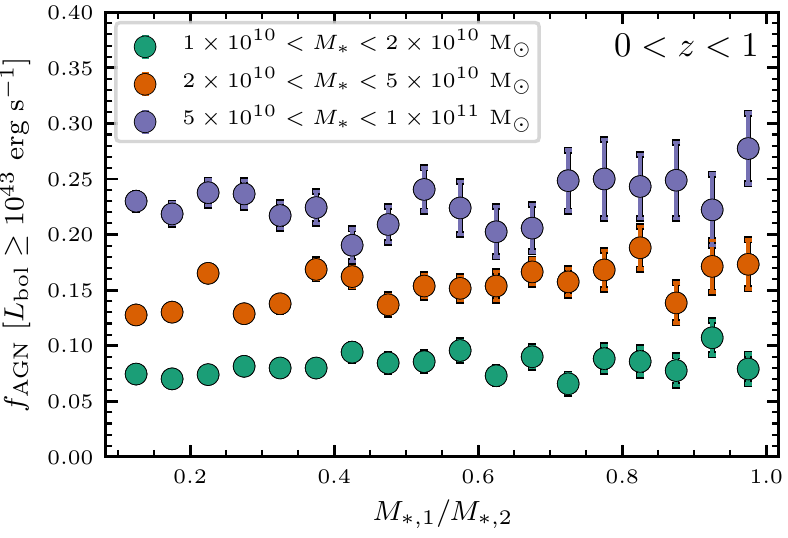}

\caption{The AGN fraction (i.e., the fraction of galaxies hosting a BH with a bolometric luminosity \Lbol $\geq 10^{43}$~~\erg) of merging galaxies in the redshift range $0<z<1$ as a function of the stellar mass ratio ($M_{\mathrm{*,1}} / M_{\mathrm{*,2}}$, where $M_{\mathrm{*,2}}$ is always the most massive member of the galaxy pair). The galaxies are split into three stellar mass ranges as indicated by the legend, and the error bars indicate the Poisson uncertainty. We find that the AGN fraction of merging galaxies is insensitive to the stellar mass ratio over the range $\frac{1}{10} < M_{\mathrm{*,1}} / M_{\mathrm{*,2}} < 1$, which is why we find similar results for this study when considering either minor mergers ($\frac{1}{10} < M_{\mathrm{*,1}} / M_{\mathrm{*,2}} < \frac{1}{4}$) or major mergers ($M_{\mathrm{*,1}} / M_{\mathrm{*,2}} \geq \frac{1}{4}$).}

\label{fig:merger_ratio}
\end{figure}

This study has focused exclusively on the influence of major mergers (i.e., $M_{\mathrm{*,1}} / M_{\mathrm{*,2}} \geq \frac{1}{4}$) as triggering mechanisms for increased BH activity. However, minor mergers may also play a role, and, as they are more common than major mergers, their importance could potentially be much larger. Here we define minor mergers as those with stellar mass ratios of $\frac{1}{10} < M_{\mathrm{*,1}} / M_{\mathrm{*,2}} < \frac{1}{4}$.

If we repeat the analysis of this study now for minor mergers we find a very similar overall result: the enhancements of the merger fractions in \cref{fig:major_merger_fraction}, the AGN fractions in \cref{fig:agn_fraction,fig:close_pairs}, and the contribution to the CBHAR in \cref{fig:cbhard} are all virtually unchanged. We can see why this is from \cref{fig:merger_ratio}, which shows the AGN fraction of merging galaxies in the redshift range $0<z<1$ as a function of the stellar mass ratio ($M_{\mathrm{*,1}} / M_{\mathrm{*,2}}$). Whilst one may have expected an increasing influence upon BH activity with increasing stellar mass ratio, instead, over the stellar mass ratio range $\frac{1}{10} < M_{\mathrm{*,1}} / M_{\mathrm{*,2}} < 1$ the AGN fraction of galaxies does not evolve.       

Therefore the conclusions we have reported for major mergers also hold true for minor mergers, in that they do enhance BH activity, yet this enhancement is not statistically meaningful for BH growth. 

\section{Conclusions}
\label{sect:conclusions}

Using the cosmological hydrodynamical \eagle simulation, we have investigated to what degree black hole (BH) activity is enhanced during the period of a major merger (i.e., those with a stellar mass ratio of $M_{\mathrm{*,1}} / M_{\mathrm{*,2}} \geq \frac{1}{4}$, where $M_{\mathrm{*,2}}$ is the most massive of the two galaxies). For this study, an \squotes{active} BH (or AGN) is defined to be one that has a bolometric AGN luminosity greater than \Lbol $\geq 10^{43}$~\erg or an Eddington rate greater than \edd $\geq 10^{-2}$ (\squotes{inactive} galaxies are therefore those with BHs accreting at rates lower than these limits, i.e., \Lbol $< 10^{43}$~\erg or \edd $< 10^{-2}$). When referring to an \squotes{excess} value below, we are referring to the ratio of two merger or AGN fractions (between the merger or AGN fractions of the selected samples and their associated control samples). 

Our main conclusions are as follows:

\begin{itemize}
    \item \textbf{AGN have a higher major merger fraction than their inactive galaxy counterparts.} The excess between the major merger fraction of AGN relative to the major merger fraction of inactive galaxies increases with increasing AGN luminosity and Eddington rate: reaching a factor of $\approx 1.75$ at \Lbol $\sim 10^{4}$~\erg, and a factor of $\approx 3$ at the Eddington limit (see \cref{fig:major_merger_fraction}).
    
    \item \textbf{There AGN fraction of major mergers is higher than the AGN fraction of their isolated galaxy counterparts.} The excess between the AGN fraction of merging and isolated systems, defined by either a cut in the AGN luminosity or Eddington rate, increases with decreasing stellar mass at $z<1$ (up to a maximum value of $\approx 1.8$ at $M_* \approx 10^{10}$~\Msol). At higher redshifts ($z>1$), the excess in the AGN fraction remains approximately constant for all stellar masses (with a value of $\approx 1.1$--1.4, see \cref{fig:agn_fraction}).   
    
    \item \textbf{The AGN fraction of galaxies with close major companions is higher than the AGN fraction of their isolated galaxy counterparts}. When an AGN is defined by a cut in the bolometric luminosity, there is a potential slight excess between the AGN fraction of galaxies with close major companions and isolated systems, oscillating around a value of $\approx 1.1$ for 3D separations lower than $r_{\mathrm{sep[Major]}} \lesssim 80$~pkpc. However, when an AGN is defined by a cut in the Eddington rate, a strong trend of an increasing excess with decreasing 3D separation is found for galaxies at $z<2$, starting at 3D separations of $50 \leq r_{\mathrm{sep[Major]}} \leq 100$~pkpc, and rising to an excess value of 1.2--1.3 at 3D separations of $\approx 10$~pkpc (see \cref{fig:close_pairs,fig:ellison_compare}).
    
    \item \textbf{The galaxies hosting the BHs with the greatest enhancement of BH activity due to a major merger are almost exclusively lower mass ($M_* \lesssim 10^{11}$~\Msol).} We find little to no enhancement of BH activity in massive ($M_* \gtrsim 10^{11}$~\Msol) active or merging systems relative to their inactive or isolated counterparts. In addition, the galaxies with the largest excess in their merger and AGN fractions above the control samples are those with higher gas fractions ($f_{\mathrm{gas}} \gtrsim 0.2$), less massive BHs ($M_{\mathrm{BH}} \lesssim 10^{7}$~\Msol) and those that are central galaxies (see \cref{fig:agn_fraction,fig:fmerger_fractions_bymass,fig:closepairs_fractions_bymass,fig:fractions_byproperty}).
    
    \item \textbf{The majority of BH activity triggered via a major merger resides within the early \emph{remnants} of merging systems}. At higher redshifts ($z>1$), $\approx 50$\% of the BH activity triggered via a major merger occurs during the dynamical time after the two galaxies have already coalesced. At lower redshifts ($z<1$), this fraction raises to $\approx 65$--75\%. In addition, at lower redshifts ($z<1$) the peak of triggered BH activity occurs $\approx 0.25$ dynamical times ($\approx 300$~Myr at $z=0.5$) after the coalescence of the two galaxies, suggesting that there is typically a significant delay between the coalescence of the two galaxies and triggered BH activity (see \cref{fig:excess_vs_ndyn_broad,fig:excess_vs_ndyn}).  
    
    \item \textbf{The excess values of both the merger fraction of AGN and the AGN fraction of merging systems increases with decreasing redshift.} Throughout our analysis we have consistently found higher excess values between the merger fraction of AGN and inactive galaxies and between the AGN fraction of merging systems and isolated galaxies with decreasing redshift. This suggests that mergers are becoming increasingly important for triggering BH activity as the universe evolves (see \cref{fig:major_merger_fraction,fig:agn_fraction,fig:close_pairs,fig:excess_vs_ndyn_broad,fig:excess_vs_ndyn}). However, the \emph{abundance} of merging systems does substantially decrease with decreasing redshift (see \cref{fig:cbhard}). 

\item \textbf{Overall, mergers are not statistically relevant fuelling mechanisms for BHs.} Whilst we have repeatably found that mergers are enhancing the amount of AGN activity within the \eagle simulation, we would argue that major (or minor see \cref{sect:minor_mergers}) mergers, as triggering mechanisms, do not contribute a significant amount to BH growth globally. Both at higher redshifts ($z \gtrsim 2$), when the majority of galaxies more massive than $M_* \geq 10^{10}$~\Msol are currently experiencing a merger, and at lower redshifts ($z \lesssim 2$), when major mergers have become a small minority of the galaxy population, merging systems typically have accretion rates that are on average 2--3 times greater than their isolated counterparts (see \cref{fig:cbhard}). However, either this level of enhancement is too small, or the time spent within major mergers is too short, to have a meaningful impact upon the final day BH mass. The BHs at the present day have, on average, accumulated the majority of their mass outwith the period(s) of a major merger (see \cref{fig:fraction_in_mergers}). Indeed, we estimate that BHs in a universe where major interactions did not enhance BH activity would have BH masses that were $\gtrsim 85$\% of the mass of BHs in the true \eagle universe (see \cref{fig:fraction_in_mergers}). \emph{Therefore it appears that the enhanced growth triggered via a merger is not a necessary component for global BH growth, and the BH population would potentially be very similar in a Universe that was absent of this enhancement.} However this does not rule out the importance, or necessity, for mergers to trigger unique events in a BHs lifetime, such as for the creation of the most highly-luminous quasars, or for initiating the rapid growth phase of BHs \citep{McAlpine2018}.        

\end{itemize}

\section*{Acknowledgements}

We thank the referee for their comments that have improved the quality of this work. This work was supported by the Academy of Finland (grant number 314238). P.H.J. acknowledges the support by the European Research Council via ERC Consolidator Grant KETJU (no. 818930). DRP and SLE gratefully acknowledge NSERC for Discovery Grants which helped to fund this research. 

This work used the DiRAC@Durham facility managed by the Institute for Computational Cosmology on behalf of the STFC DiRAC HPC Facility (\url{www.dirac.ac.uk}). The equipment was funded by BEIS capital funding via STFC capital grants ST/K00042X/1, ST/P002293/1, ST/R002371/1 and ST/S002502/1, Durham University and STFC operations grant ST/R000832/1. DiRAC is part of the National e-Infrastructure.

Funding for the SDSS and SDSS-II has been provided by the Alfred P. Sloan Foundation, the Participating Institutions, the National Science Foundation, the U.S. Department of Energy, the National Aeronautics and Space Administration, the Japanese Monbukagakusho, the Max Planck Society, and the Higher Education Funding Council for England. The SDSS Web Site is \url{http://www.sdss.org/}.

The SDSS is managed by the Astrophysical Research Consortium for the Participating Institutions. The Participating Institutions are the American Museum of Natural History, Astrophysical Institute Potsdam, University of Basel, University of Cambridge, Case Western Reserve University, University of Chicago, Drexel University, Fermilab, the Institute for Advanced Study, the Japan Participation Group, Johns Hopkins University, the Joint Institute for Nuclear Astrophysics, the Kavli Institute for Particle Astrophysics and Cosmology, the Korean Scientist Group, the Chinese Academy of Sciences (LAMOST), Los Alamos National Laboratory, the Max-Planck-Institute for Astronomy (MPIA), the Max-Planck-Institute for Astrophysics (MPA), New Mexico State University, Ohio State University, University of Pittsburgh, University of Portsmouth, Princeton University, the United States Naval Observatory, and the University of Washington.

\bibliographystyle{mnras}
\bibliography{mybibfile}

\begin{thebibliography}{}
\makeatletter
\relax
\def\mn@urlcharsother{\let\do\@makeother \do\$\do\&\do\#\do\^\do\_\do\%\do\~}
\def\mn@doi{\begingroup\mn@urlcharsother \@ifnextchar [ {\mn@doi@}
  {\mn@doi@[]}}
\def\mn@doi@[#1]#2{\def\@tempa{#1}\ifx\@tempa\@empty \href
  {http://dx.doi.org/#2} {doi:#2}\else \href {http://dx.doi.org/#2} {#1}\fi
  \endgroup}
\def\mn@eprint#1#2{\mn@eprint@#1:#2::\@nil}
\def\mn@eprint@arXiv#1{\href {http://arxiv.org/abs/#1} {{\tt arXiv:#1}}}
\def\mn@eprint@dblp#1{\href {http://dblp.uni-trier.de/rec/bibtex/#1.xml}
  {dblp:#1}}
\def\mn@eprint@#1:#2:#3:#4\@nil{\def\@tempa {#1}\def\@tempb {#2}\def\@tempc
  {#3}\ifx \@tempc \@empty \let \@tempc \@tempb \let \@tempb \@tempa \fi \ifx
  \@tempb \@empty \def\@tempb {arXiv}\fi \@ifundefined
  {mn@eprint@\@tempb}{\@tempb:\@tempc}{\expandafter \expandafter \csname
  mn@eprint@\@tempb\endcsname \expandafter{\@tempc}}}

\bibitem[\protect\citeauthoryear{{Barnes} \& {Hernquist}}{{Barnes} \&
  {Hernquist}}{1991}]{BarnesAndHernquist1991}
{Barnes} J.~E.,  {Hernquist} L.~E.,  1991, \mn@doi [\apjl] {10.1086/185978},
  \href {http://adsabs.harvard.edu/abs/1991ApJ...370L..65B} {370, L65}

\bibitem[\protect\citeauthoryear{{Blumenthal} \& {Barnes}}{{Blumenthal} \&
  {Barnes}}{2018}]{Blumenthal2018}
{Blumenthal} K.~A.,  {Barnes} J.~E.,  2018, \mn@doi [\mnras]
  {10.1093/mnras/sty1605}, \href
  {https://ui.adsabs.harvard.edu/abs/2018MNRAS.479.3952B} {479, 3952}

\bibitem[\protect\citeauthoryear{{Bondi} \& {Hoyle}}{{Bondi} \&
  {Hoyle}}{1944}]{Bondi1944}
{Bondi} H.,  {Hoyle} F.,  1944, \mnras, \href
  {http://adsabs.harvard.edu/abs/1944MNRAS.104..273B} {104, 273}

\bibitem[\protect\citeauthoryear{{Bottrell} et~al.,}{{Bottrell}
  et~al.}{2019}]{Bottrell2019}
{Bottrell} C.,  et~al., 2019, \mn@doi [\mnras] {10.1093/mnras/stz2934}, \href
  {https://ui.adsabs.harvard.edu/abs/2019MNRAS.490.5390B} {490, 5390}

\bibitem[\protect\citeauthoryear{{Brandt} \& {Alexander}}{{Brandt} \&
  {Alexander}}{2015}]{Brandt2015}
{Brandt} W.~N.,  {Alexander} D.~M.,  2015, \mn@doi [\aapr]
  {10.1007/s00159-014-0081-z}, \href
  {https://ui.adsabs.harvard.edu/abs/2015A&ARv..23....1B} {23, 1}

\bibitem[\protect\citeauthoryear{{Chabrier}}{{Chabrier}}{2003}]{Chabrier2003}
{Chabrier} G.,  2003, \mn@doi [\pasp] {10.1086/376392}, \href
  {http://adsabs.harvard.edu/abs/2003PASP..115..763C} {115, 763}

\bibitem[\protect\citeauthoryear{{Cisternas} et~al.,}{{Cisternas}
  et~al.}{2011}]{Cisternas2011}
{Cisternas} M.,  et~al., 2011, \mn@doi [\apj] {10.1088/0004-637X/726/2/57},
  \href {http://adsabs.harvard.edu/abs/2011ApJ...726...57C} {726, 57}

\bibitem[\protect\citeauthoryear{{Cotini}, {Ripamonti}, {Caccianiga}, {Colpi},
  {Della Ceca}, {Mapelli}, {Severgnini}  \& {Segreto}}{{Cotini}
  et~al.}{2013}]{Cotini2013}
{Cotini} S.,  {Ripamonti} E.,  {Caccianiga} A.,  {Colpi} M.,  {Della Ceca} R.,
  {Mapelli} M.,  {Severgnini} P.,   {Segreto} A.,  2013, \mn@doi [\mnras]
  {10.1093/mnras/stt358}, \href
  {http://adsabs.harvard.edu/abs/2013MNRAS.431.2661C} {431, 2661}

\bibitem[\protect\citeauthoryear{{Crain} et~al.,}{{Crain}
  et~al.}{2015}]{Crain2015}
{Crain} R.~A.,  et~al., 2015, \mn@doi [\mnras] {10.1093/mnras/stv725}, \href
  {http://adsabs.harvard.edu/abs/2015MNRAS.450.1937C} {450, 1937}

\bibitem[\protect\citeauthoryear{{Dalla Vecchia} \& {Schaye}}{{Dalla Vecchia}
  \& {Schaye}}{2012}]{DallaVecchia_Schaye2012}
{Dalla Vecchia} C.,  {Schaye} J.,  2012, \mn@doi [\mnras]
  {10.1111/j.1365-2966.2012.21704.x}, \href
  {http://adsabs.harvard.edu/abs/2012MNRAS.426..140D} {426, 140}

\bibitem[\protect\citeauthoryear{{Di Matteo}, {Springel}  \& {Hernquist}}{{Di
  Matteo} et~al.}{2005}]{DiMatteo2005}
{Di Matteo} T.,  {Springel} V.,   {Hernquist} L.,  2005, \mn@doi [\nat]
  {10.1038/nature03335}, \href
  {http://adsabs.harvard.edu/abs/2005Natur.433..604D} {433, 604}

\bibitem[\protect\citeauthoryear{{Dolag}, {Borgani}, {Murante}  \&
  {Springel}}{{Dolag} et~al.}{2009}]{Dolag2009}
{Dolag} K.,  {Borgani} S.,  {Murante} G.,   {Springel} V.,  2009, \mn@doi
  [\mnras] {10.1111/j.1365-2966.2009.15034.x}, \href
  {http://adsabs.harvard.edu/abs/2009MNRAS.399..497D} {399, 497}

\bibitem[\protect\citeauthoryear{{Dubois}, {Volonteri}, {Silk}, {Devriendt},
  {Slyz}  \& {Teyssier}}{{Dubois} et~al.}{2015}]{Dubois2015}
{Dubois} Y.,  {Volonteri} M.,  {Silk} J.,  {Devriendt} J.,  {Slyz} A.,
  {Teyssier} R.,  2015, \mn@doi [\mnras] {10.1093/mnras/stv1416}, \href
  {http://adsabs.harvard.edu/abs/2015MNRAS.452.1502D} {452, 1502}

\bibitem[\protect\citeauthoryear{{Ellison}, {Patton}, {Simard}, {McConnachie},
  {Baldry}  \& {Mendel}}{{Ellison} et~al.}{2010}]{Ellison2010}
{Ellison} S.~L.,  {Patton} D.~R.,  {Simard} L.,  {McConnachie} A.~W.,  {Baldry}
  I.~K.,   {Mendel} J.~T.,  2010, \mn@doi [\mnras]
  {10.1111/j.1365-2966.2010.17076.x}, \href
  {https://ui.adsabs.harvard.edu/abs/2010MNRAS.407.1514E} {407, 1514}

\bibitem[\protect\citeauthoryear{{Ellison}, {Patton}, {Mendel}  \&
  {Scudder}}{{Ellison} et~al.}{2011}]{Ellison2011}
{Ellison} S.~L.,  {Patton} D.~R.,  {Mendel} J.~T.,   {Scudder} J.~M.,  2011,
  \mn@doi [\mnras] {10.1111/j.1365-2966.2011.19624.x}, \href
  {https://ui.adsabs.harvard.edu/abs/2011MNRAS.418.2043E} {418, 2043}

\bibitem[\protect\citeauthoryear{{Ellison}, {Mendel}, {Patton}  \&
  {Scudder}}{{Ellison} et~al.}{2013}]{Ellison2013}
{Ellison} S.~L.,  {Mendel} J.~T.,  {Patton} D.~R.,   {Scudder} J.~M.,  2013,
  \mn@doi [\mnras] {10.1093/mnras/stt1562}, \href
  {http://adsabs.harvard.edu/abs/2013MNRAS.435.3627E} {435, 3627}

\bibitem[\protect\citeauthoryear{{Ellison}, {Patton}  \& {Hickox}}{{Ellison}
  et~al.}{2015}]{Ellison2015}
{Ellison} S.~L.,  {Patton} D.~R.,   {Hickox} R.~C.,  2015, \mn@doi [\mnras]
  {10.1093/mnrasl/slv061}, \href
  {https://ui.adsabs.harvard.edu/abs/2015MNRAS.451L..35E} {451, L35}

\bibitem[\protect\citeauthoryear{{Ellison}, {Viswanathan}, {Patton},
  {Bottrell}, {McConnachie}, {Gwyn}  \& {Cuillandre}}{{Ellison}
  et~al.}{2019}]{Ellison2019}
{Ellison} S.~L.,  {Viswanathan} A.,  {Patton} D.~R.,  {Bottrell} C.,
  {McConnachie} A.~W.,  {Gwyn} S.,   {Cuillandre} J.-C.,  2019, \mn@doi
  [\mnras] {10.1093/mnras/stz1431}, \href
  {https://ui.adsabs.harvard.edu/abs/2019MNRAS.487.2491E} {487, 2491}

\bibitem[\protect\citeauthoryear{{Fan} et~al.,}{{Fan} et~al.}{2016}]{Fan2016}
{Fan} L.,  et~al., 2016, \mn@doi [\apjl] {10.3847/2041-8205/822/2/L32}, \href
  {http://adsabs.harvard.edu/abs/2016ApJ...822L..32F} {822, L32}

\bibitem[\protect\citeauthoryear{{Furlong} et~al.,}{{Furlong}
  et~al.}{2015}]{Furlong2015}
{Furlong} M.,  et~al., 2015, \mn@doi [\mnras] {10.1093/mnras/stv852}, \href
  {http://adsabs.harvard.edu/abs/2015MNRAS.450.4486F} {450, 4486}

\bibitem[\protect\citeauthoryear{{Furlong} et~al.,}{{Furlong}
  et~al.}{2017}]{Furlong2017}
{Furlong} M.,  et~al., 2017, \mn@doi [\mnras] {10.1093/mnras/stw2740}, \href
  {http://adsabs.harvard.edu/abs/2017MNRAS.465..722F} {465, 722}

\bibitem[\protect\citeauthoryear{{Glikman}, {Simmons}, {Mailly}, {Schawinski},
  {Urry}  \& {Lacy}}{{Glikman} et~al.}{2015}]{Glikman2015}
{Glikman} E.,  {Simmons} B.,  {Mailly} M.,  {Schawinski} K.,  {Urry} C.~M.,
  {Lacy} M.,  2015, \mn@doi [\apj] {10.1088/0004-637X/806/2/218}, \href
  {http://adsabs.harvard.edu/abs/2015ApJ...806..218G} {806, 218}

\bibitem[\protect\citeauthoryear{{Goulding} \& {Alexander}}{{Goulding} \&
  {Alexander}}{2009}]{Goulding2009}
{Goulding} A.~D.,  {Alexander} D.~M.,  2009, \mn@doi [\mnras]
  {10.1111/j.1365-2966.2009.15194.x}, \href
  {http://adsabs.harvard.edu/abs/2009MNRAS.398.1165G} {398, 1165}

\bibitem[\protect\citeauthoryear{{Goulding} et~al.,}{{Goulding}
  et~al.}{2018}]{Goulding2018}
{Goulding} A.~D.,  et~al., 2018, \mn@doi [\pasj] {10.1093/pasj/psx135}, \href
  {https://ui.adsabs.harvard.edu/abs/2018PASJ...70S..37G} {70, S37}

\bibitem[\protect\citeauthoryear{{Hewlett}, {Villforth}, {Wild},
  {Mendez-Abreu}, {Pawlik}  \& {Rowlands}}{{Hewlett}
  et~al.}{2017}]{Hewlett2017}
{Hewlett} T.,  {Villforth} C.,  {Wild} V.,  {Mendez-Abreu} J.,  {Pawlik} M.,
  {Rowlands} K.,  2017, \mn@doi [\mnras] {10.1093/mnras/stx997}, \href
  {https://ui.adsabs.harvard.edu/abs/2017MNRAS.470..755H} {470, 755}

\bibitem[\protect\citeauthoryear{{Hickox}, {Mullaney}, {Alexander}, {Chen},
  {Civano}, {Goulding}  \& {Hainline}}{{Hickox} et~al.}{2014}]{Hickox2014}
{Hickox} R.~C.,  {Mullaney} J.~R.,  {Alexander} D.~M.,  {Chen} C.-T.~J.,
  {Civano} F.~M.,  {Goulding} A.~D.,   {Hainline} K.~N.,  2014, \mn@doi [\apj]
  {10.1088/0004-637X/782/1/9}, \href
  {http://adsabs.harvard.edu/abs/2014ApJ...782....9H} {782, 9}

\bibitem[\protect\citeauthoryear{{Hirschmann}, {Khochfar}, {Burkert}, {Naab},
  {Genel}  \& {Somerville}}{{Hirschmann} et~al.}{2010}]{Hirschmann2010}
{Hirschmann} M.,  {Khochfar} S.,  {Burkert} A.,  {Naab} T.,  {Genel} S.,
  {Somerville} R.~S.,  2010, \mn@doi [\mnras]
  {10.1111/j.1365-2966.2010.17006.x}, \href
  {http://adsabs.harvard.edu/abs/2010MNRAS.407.1016H} {407, 1016}

\bibitem[\protect\citeauthoryear{{Hirschmann}, {Dolag}, {Saro}, {Bachmann},
  {Borgani}  \& {Burkert}}{{Hirschmann} et~al.}{2014}]{Hirschmann2014}
{Hirschmann} M.,  {Dolag} K.,  {Saro} A.,  {Bachmann} L.,  {Borgani} S.,
  {Burkert} A.,  2014, \mn@doi [\mnras] {10.1093/mnras/stu1023}, \href
  {https://ui.adsabs.harvard.edu/abs/2014MNRAS.442.2304H} {442, 2304}

\bibitem[\protect\citeauthoryear{{Hopkins}, {Hernquist}, {Cox}  \& {Kere{\v
  s}}}{{Hopkins} et~al.}{2008}]{Hopkins2008}
{Hopkins} P.~F.,  {Hernquist} L.,  {Cox} T.~J.,   {Kere{\v s}} D.,  2008,
  \mn@doi [\apjs] {10.1086/524362}, \href
  {http://adsabs.harvard.edu/abs/2008ApJS..175..356H} {175, 356}

\bibitem[\protect\citeauthoryear{{Jahnke} \& {Macci{\`o}}}{{Jahnke} \&
  {Macci{\`o}}}{2011}]{Jahnke2011}
{Jahnke} K.,  {Macci{\`o}} A.~V.,  2011, \mn@doi [\apj]
  {10.1088/0004-637X/734/2/92}, \href
  {http://adsabs.harvard.edu/abs/2011ApJ...734...92J} {734, 92}

\bibitem[\protect\citeauthoryear{{Ji}, {Peirani}  \& {Yi}}{{Ji}
  et~al.}{2014}]{Ji2014}
{Ji} I.,  {Peirani} S.,   {Yi} S.~K.,  2014, \mn@doi [\aap]
  {10.1051/0004-6361/201423530}, \href
  {https://ui.adsabs.harvard.edu/abs/2014A&A...566A..97J} {566, A97}

\bibitem[\protect\citeauthoryear{{Johansson}, {Burkert}  \& {Naab}}{{Johansson}
  et~al.}{2009}]{Johansson2009}
{Johansson} P.~H.,  {Burkert} A.,   {Naab} T.,  2009, \mn@doi [\apjl]
  {10.1088/0004-637X/707/2/L184}, \href
  {https://ui.adsabs.harvard.edu/abs/2009ApJ...707L.184J} {707, L184}

\bibitem[\protect\citeauthoryear{{Kauffmann} et~al.,}{{Kauffmann}
  et~al.}{2003}]{Kauffmann2003}
{Kauffmann} G.,  et~al., 2003, \mn@doi [\mnras]
  {10.1046/j.1365-8711.2003.06291.x}, \href
  {https://ui.adsabs.harvard.edu/abs/2003MNRAS.341...33K} {341, 33}

\bibitem[\protect\citeauthoryear{{Kocevski} et~al.,}{{Kocevski}
  et~al.}{2012}]{Kocevski2012}
{Kocevski} D.~D.,  et~al., 2012, \mn@doi [\apj] {10.1088/0004-637X/744/2/148},
  \href {http://adsabs.harvard.edu/abs/2012ApJ...744..148K} {744, 148}

\bibitem[\protect\citeauthoryear{{Koss}, {Mushotzky}, {Veilleux}  \&
  {Winter}}{{Koss} et~al.}{2010}]{Koss2010}
{Koss} M.,  {Mushotzky} R.,  {Veilleux} S.,   {Winter} L.,  2010, \mn@doi
  [\apjl] {10.1088/2041-8205/716/2/L125}, \href
  {http://adsabs.harvard.edu/abs/2010ApJ...716L.125K} {716, L125}

\bibitem[\protect\citeauthoryear{{Koss} et~al.,}{{Koss}
  et~al.}{2018}]{Koss2018}
{Koss} M.~J.,  et~al., 2018, \mn@doi [\nat] {10.1038/s41586-018-0652-7}, \href
  {https://ui.adsabs.harvard.edu/abs/2018Natur.563..214K} {563, 214}

\bibitem[\protect\citeauthoryear{{Lah{\'e}n}, {Johansson}, {Rantala}, {Naab}
  \& {Frigo}}{{Lah{\'e}n} et~al.}{2018}]{Lahen2018}
{Lah{\'e}n} N.,  {Johansson} P.~H.,  {Rantala} A.,  {Naab} T.,   {Frigo} M.,
  2018, \mn@doi [\mnras] {10.1093/mnras/sty060-}, \href
  {https://ui.adsabs.harvard.edu/abs/2018MNRAS.475.3934L} {475, 3934}

\bibitem[\protect\citeauthoryear{{Lotz}, {Jonsson}, {Cox}  \& {Primack}}{{Lotz}
  et~al.}{2010}]{Lotz2010}
{Lotz} J.~M.,  {Jonsson} P.,  {Cox} T.~J.,   {Primack} J.~R.,  2010, \mn@doi
  [\mnras] {10.1111/j.1365-2966.2010.16268.x}, \href
  {https://ui.adsabs.harvard.edu/abs/2010MNRAS.404..575L} {404, 575}

\bibitem[\protect\citeauthoryear{{Magorrian} et~al.,}{{Magorrian}
  et~al.}{1998}]{Magorrian1998}
{Magorrian} J.,  et~al., 1998, \mn@doi [\aj] {10.1086/300353}, \href
  {http://adsabs.harvard.edu/abs/1998AJ....115.2285M} {115, 2285}

\bibitem[\protect\citeauthoryear{{Marian} et~al.,}{{Marian}
  et~al.}{2019}]{Marian2019}
{Marian} V.,  et~al., 2019, \mn@doi [\apj] {10.3847/1538-4357/ab385b}, \href
  {https://ui.adsabs.harvard.edu/abs/2019ApJ...882..141M} {882, 141}

\bibitem[\protect\citeauthoryear{{Martin} et~al.,}{{Martin}
  et~al.}{2018}]{Martin2018}
{Martin} G.,  et~al., 2018, \mn@doi [\mnras] {10.1093/mnras/sty324}, \href
  {https://ui.adsabs.harvard.edu/abs/2018MNRAS.476.2801M} {476, 2801}

\bibitem[\protect\citeauthoryear{{McAlpine} et~al.,}{{McAlpine}
  et~al.}{2016}]{McAlpine2015}
{McAlpine} S.,  et~al., 2016, \mn@doi [Astronomy and Computing]
  {10.1016/j.ascom.2016.02.004}, \href
  {http://adsabs.harvard.edu/abs/2016A%26C....15...72M} {15, 72}

\bibitem[\protect\citeauthoryear{{McAlpine}, {Bower}, {Harrison}, {Crain},
  {Schaller}, {Schaye}  \& {Theuns}}{{McAlpine} et~al.}{2017}]{McAlpine2017}
{McAlpine} S.,  {Bower} R.~G.,  {Harrison} C.~M.,  {Crain} R.~A.,  {Schaller}
  M.,  {Schaye} J.,   {Theuns} T.,  2017, \mn@doi [\mnras]
  {10.1093/mnras/stx658}, \href
  {http://adsabs.harvard.edu/abs/2017MNRAS.468.3395M} {468, 3395}

\bibitem[\protect\citeauthoryear{{McAlpine}, {Bower}, {Rosario}, {Crain},
  {Schaye}  \& {Theuns}}{{McAlpine} et~al.}{2018}]{McAlpine2018}
{McAlpine} S.,  {Bower} R.~G.,  {Rosario} D.~J.,  {Crain} R.~A.,  {Schaye} J.,
   {Theuns} T.,  2018, \mn@doi [\mnras] {10.1093/mnras/sty2489}, \href
  {http://adsabs.harvard.edu/abs/2018MNRAS.481.3118M} {481, 3118}

\bibitem[\protect\citeauthoryear{{McConnell} \& {Ma}}{{McConnell} \&
  {Ma}}{2013}]{McConnellandMa2013}
{McConnell} N.~J.,  {Ma} C.-P.,  2013, \mn@doi [ApJ]
  {10.1088/0004-637X/764/2/184}, \href
  {http://adsabs.harvard.edu/abs/2013ApJ...764..184M} {764, 184}

\bibitem[\protect\citeauthoryear{{Mechtley} et~al.,}{{Mechtley}
  et~al.}{2016}]{Mechtley2016}
{Mechtley} M.,  et~al., 2016, \mn@doi [\apj] {10.3847/0004-637X/830/2/156},
  \href {http://adsabs.harvard.edu/abs/2016ApJ...830..156M} {830, 156}

\bibitem[\protect\citeauthoryear{{Mihos} \& {Hernquist}}{{Mihos} \&
  {Hernquist}}{1996}]{MihosAndHernquist1996}
{Mihos} J.~C.,  {Hernquist} L.,  1996, \mn@doi [\apj] {10.1086/177353}, \href
  {http://adsabs.harvard.edu/abs/1996ApJ...464..641M} {464, 641}

\bibitem[\protect\citeauthoryear{{Patton}, {Torrey}, {Ellison}, {Mendel}  \&
  {Scudder}}{{Patton} et~al.}{2013}]{Patton2013}
{Patton} D.~R.,  {Torrey} P.,  {Ellison} S.~L.,  {Mendel} J.~T.,   {Scudder}
  J.~M.,  2013, \mn@doi [\mnras] {10.1093/mnrasl/slt058}, \href
  {https://ui.adsabs.harvard.edu/abs/2013MNRAS.433L..59P} {433, L59}

\bibitem[\protect\citeauthoryear{{Patton}, {Qamar}, {Ellison}, {Bluck},
  {Simard}, {Mendel}, {Moreno}  \& {Torrey}}{{Patton}
  et~al.}{2016}]{Patton2016}
{Patton} D.~R.,  {Qamar} F.~D.,  {Ellison} S.~L.,  {Bluck} A. F.~L.,  {Simard}
  L.,  {Mendel} J.~T.,  {Moreno} J.,   {Torrey} P.,  2016, \mn@doi [\mnras]
  {10.1093/mnras/stw1494}, \href
  {https://ui.adsabs.harvard.edu/abs/2016MNRAS.461.2589P} {461, 2589}

\bibitem[\protect\citeauthoryear{{Pawlik}, {Wild}, {Walcher}, {Johansson},
  {Villforth}, {Rowlands}, {Mendez-Abreu}  \& {Hewlett}}{{Pawlik}
  et~al.}{2016}]{Pawlik2016}
{Pawlik} M.~M.,  {Wild} V.,  {Walcher} C.~J.,  {Johansson} P.~H.,  {Villforth}
  C.,  {Rowlands} K.,  {Mendez-Abreu} J.,   {Hewlett} T.,  2016, \mn@doi
  [\mnras] {10.1093/mnras/stv2878}, \href
  {https://ui.adsabs.harvard.edu/abs/2016MNRAS.456.3032P} {456, 3032}

\bibitem[\protect\citeauthoryear{{Peng}}{{Peng}}{2007}]{Peng2007}
{Peng} C.~Y.,  2007, \mn@doi [\apj] {10.1086/522774}, \href
  {http://adsabs.harvard.edu/abs/2007ApJ...671.1098P} {671, 1098}

\bibitem[\protect\citeauthoryear{{Planck Collaboration} et~al.,}{{Planck
  Collaboration} et~al.}{2014}]{Planck2013}
{Planck Collaboration} et~al., 2014, \mn@doi [Astronomy and Astrophysics]
  {10.1051/0004-6361/201321529}, \href
  {http://adsabs.harvard.edu/abs/2014A%26A...571A...1P} {571, A1}

\bibitem[\protect\citeauthoryear{{Pontzen}, {Tremmel}, {Roth}, {Peiris},
  {Saintonge}, {Volonteri}, {Quinn}  \& {Governato}}{{Pontzen}
  et~al.}{2017}]{Pontzen2017}
{Pontzen} A.,  {Tremmel} M.,  {Roth} N.,  {Peiris} H.~V.,  {Saintonge} A.,
  {Volonteri} M.,  {Quinn} T.,   {Governato} F.,  2017, \mn@doi [\mnras]
  {10.1093/mnras/stw2627}, \href
  {http://adsabs.harvard.edu/abs/2017MNRAS.465..547P} {465, 547}

\bibitem[\protect\citeauthoryear{{Qu} et~al.,}{{Qu} et~al.}{2017}]{Qu2017}
{Qu} Y.,  et~al., 2017, \mn@doi [\mnras] {10.1093/mnras/stw2437}, \href
  {http://adsabs.harvard.edu/abs/2017MNRAS.464.1659Q} {464, 1659}

\bibitem[\protect\citeauthoryear{{Rantala}, {Pihajoki}, {Johansson}, {Naab},
  {Lah{\'e}n}  \& {Sawala}}{{Rantala} et~al.}{2017}]{Rantala2017}
{Rantala} A.,  {Pihajoki} P.,  {Johansson} P.~H.,  {Naab} T.,  {Lah{\'e}n} N.,
   {Sawala} T.,  2017, \mn@doi [\apj] {10.3847/1538-4357/aa6d65}, \href
  {https://ui.adsabs.harvard.edu/abs/2017ApJ...840...53R} {840, 53}

\bibitem[\protect\citeauthoryear{{Rodighiero} et~al.,}{{Rodighiero}
  et~al.}{2015}]{Rodighiero2015}
{Rodighiero} G.,  et~al., 2015, \mn@doi [\apjl] {10.1088/2041-8205/800/1/L10},
  \href {http://adsabs.harvard.edu/abs/2015ApJ...800L..10R} {800, L10}

\bibitem[\protect\citeauthoryear{{Rodriguez-Gomez} et~al.,}{{Rodriguez-Gomez}
  et~al.}{2015}]{Rodriguez-Gomez2015}
{Rodriguez-Gomez} V.,  et~al., 2015, \mn@doi [\mnras] {10.1093/mnras/stv264},
  \href {http://adsabs.harvard.edu/abs/2015MNRAS.449...49R} {449, 49}

\bibitem[\protect\citeauthoryear{{Rodr{\'\i}guez Montero}, {Dav{\'e}}, {Wild},
  {Angl{\'e}s-Alc{\'a}zar}  \& {Narayanan}}{{Rodr{\'\i}guez Montero}
  et~al.}{2019}]{RodriguezMontero2019}
{Rodr{\'\i}guez Montero} F.,  {Dav{\'e}} R.,  {Wild} V.,
  {Angl{\'e}s-Alc{\'a}zar} D.,   {Narayanan} D.,  2019, \mn@doi [\mnras]
  {10.1093/mnras/stz2580}, \href
  {https://ui.adsabs.harvard.edu/abs/2019MNRAS.490.2139R} {490, 2139}

\bibitem[\protect\citeauthoryear{{Rosario} et~al.,}{{Rosario}
  et~al.}{2015}]{Rosario2015}
{Rosario} D.~J.,  et~al., 2015, \mn@doi [\aap] {10.1051/0004-6361/201423782},
  \href {http://adsabs.harvard.edu/abs/2015A%26A...573A..85R} {573, A85}

\bibitem[\protect\citeauthoryear{{Rosas-Guevara} et~al.,}{{Rosas-Guevara}
  et~al.}{2015}]{RosasGuevara2015}
{Rosas-Guevara} Y.~M.,  et~al., 2015, \mn@doi [\mnras] {10.1093/mnras/stv2056},
  \href {http://adsabs.harvard.edu/abs/2015MNRAS.454.1038R} {454, 1038}

\bibitem[\protect\citeauthoryear{{Rosas-Guevara}, {Bower}, {Schaye},
  {McAlpine}, {Dalla Vecchia}, {Frenk}, {Schaller}  \&
  {Theuns}}{{Rosas-Guevara} et~al.}{2016}]{RosasGuevara2016}
{Rosas-Guevara} Y.,  {Bower} R.~G.,  {Schaye} J.,  {McAlpine} S.,  {Dalla
  Vecchia} C.,  {Frenk} C.~S.,  {Schaller} M.,   {Theuns} T.,  2016, \mn@doi
  [\mnras] {10.1093/mnras/stw1679}, \href
  {http://adsabs.harvard.edu/abs/2016MNRAS.462..190R} {462, 190}

\bibitem[\protect\citeauthoryear{{Salcido}, {Bower}, {Theuns}, {McAlpine},
  {Schaller}, {Crain}, {Schaye}  \& {Regan}}{{Salcido}
  et~al.}{2016}]{Salcido2016}
{Salcido} J.,  {Bower} R.~G.,  {Theuns} T.,  {McAlpine} S.,  {Schaller} M.,
  {Crain} R.~A.,  {Schaye} J.,   {Regan} J.,  2016, \mn@doi [\mnras]
  {10.1093/mnras/stw2048}, \href
  {https://ui.adsabs.harvard.edu/abs/2016MNRAS.463..870S} {463, 870}

\bibitem[\protect\citeauthoryear{{Sanders}, {Soifer}, {Elias}, {Madore},
  {Matthews}, {Neugebauer}  \& {Scoville}}{{Sanders}
  et~al.}{1988}]{Sanders1988}
{Sanders} D.~B.,  {Soifer} B.~T.,  {Elias} J.~H.,  {Madore} B.~F.,  {Matthews}
  K.,  {Neugebauer} G.,   {Scoville} N.~Z.,  1988, \mn@doi [\apj]
  {10.1086/165983}, \href {http://adsabs.harvard.edu/abs/1988ApJ...325...74S}
  {325, 74}

\bibitem[\protect\citeauthoryear{{Satyapal}, {Ellison}, {McAlpine}, {Hickox},
  {Patton}  \& {Mendel}}{{Satyapal} et~al.}{2014}]{Satyapal2014}
{Satyapal} S.,  {Ellison} S.~L.,  {McAlpine} W.,  {Hickox} R.~C.,  {Patton}
  D.~R.,   {Mendel} J.~T.,  2014, \mn@doi [\mnras] {10.1093/mnras/stu650},
  \href {http://adsabs.harvard.edu/abs/2014MNRAS.441.1297S} {441, 1297}

\bibitem[\protect\citeauthoryear{{Schawinski}, {Treister}, {Urry}, {Cardamone},
  {Simmons}  \& {Yi}}{{Schawinski} et~al.}{2011}]{Schawinski2011}
{Schawinski} K.,  {Treister} E.,  {Urry} C.~M.,  {Cardamone} C.~N.,  {Simmons}
  B.,   {Yi} S.~K.,  2011, \mn@doi [\apjl] {10.1088/2041-8205/727/2/L31}, \href
  {http://adsabs.harvard.edu/abs/2011ApJ...727L..31S} {727, L31}

\bibitem[\protect\citeauthoryear{{Schawinski}, {Simmons}, {Urry}, {Treister}
  \& {Glikman}}{{Schawinski} et~al.}{2012}]{Schawinski2012}
{Schawinski} K.,  {Simmons} B.~D.,  {Urry} C.~M.,  {Treister} E.,   {Glikman}
  E.,  2012, \mn@doi [\mnras] {10.1111/j.1745-3933.2012.01302.x}, \href
  {http://adsabs.harvard.edu/abs/2012MNRAS.425L..61S} {425, L61}

\bibitem[\protect\citeauthoryear{{Schawinski}, {Koss}, {Berney}  \&
  {Sartori}}{{Schawinski} et~al.}{2015}]{Schawinski2015}
{Schawinski} K.,  {Koss} M.,  {Berney} S.,   {Sartori} L.~F.,  2015, \mn@doi
  [\mnras] {10.1093/mnras/stv1136}, \href
  {http://adsabs.harvard.edu/abs/2015MNRAS.451.2517S} {451, 2517}

\bibitem[\protect\citeauthoryear{{Schaye} \& {Dalla Vecchia}}{{Schaye} \&
  {Dalla Vecchia}}{2008}]{Schaye2008}
{Schaye} J.,  {Dalla Vecchia} C.,  2008, \mn@doi [\mnras]
  {10.1111/j.1365-2966.2007.12639.x}, \href
  {http://adsabs.harvard.edu/abs/2008MNRAS.383.1210S} {383, 1210}

\bibitem[\protect\citeauthoryear{{Schaye} et~al.,}{{Schaye}
  et~al.}{2015}]{Schaye2015}
{Schaye} J.,  et~al., 2015, \mn@doi [\mnras] {10.1093/mnras/stu2058}, \href
  {http://adsabs.harvard.edu/abs/2015MNRAS.446..521S} {446, 521}

\bibitem[\protect\citeauthoryear{{Scholtz} et~al.,}{{Scholtz}
  et~al.}{2018}]{Scholtz2018}
{Scholtz} J.,  et~al., 2018, \mn@doi [\mnras] {10.1093/mnras/stx3177}, \href
  {https://ui.adsabs.harvard.edu/abs/2018MNRAS.475.1288S} {475, 1288}

\bibitem[\protect\citeauthoryear{{Shakura} \& {Sunyaev}}{{Shakura} \&
  {Sunyaev}}{1973}]{Shakura1973}
{Shakura} N.~I.,  {Sunyaev} R.~A.,  1973, \aap, \href
  {http://adsabs.harvard.edu/abs/1973A%26A....24..337S} {24, 337}

\bibitem[\protect\citeauthoryear{{Silverman} et~al.,}{{Silverman}
  et~al.}{2011}]{Silverman2011}
{Silverman} J.~D.,  et~al., 2011, \mn@doi [\apj] {10.1088/0004-637X/743/1/2},
  \href {https://ui.adsabs.harvard.edu/abs/2011ApJ...743....2S} {743, 2}

\bibitem[\protect\citeauthoryear{{Snyder}, {Rodriguez-Gomez}, {Lotz}, {Torrey},
  {Quirk}, {Hernquist}, {Vogelsberger}  \& {Freeman}}{{Snyder}
  et~al.}{2019}]{Snyder2019}
{Snyder} G.~F.,  {Rodriguez-Gomez} V.,  {Lotz} J.~M.,  {Torrey} P.,  {Quirk} A.
  C.~N.,  {Hernquist} L.,  {Vogelsberger} M.,   {Freeman} P.~E.,  2019, \mn@doi
  [\mnras] {10.1093/mnras/stz1059}, \href
  {https://ui.adsabs.harvard.edu/abs/2019MNRAS.486.3702S} {486, 3702}

\bibitem[\protect\citeauthoryear{{Springel}, {White}, {Tormen}  \&
  {Kauffmann}}{{Springel} et~al.}{2001}]{Springel2001}
{Springel} V.,  {White} S.~D.~M.,  {Tormen} G.,   {Kauffmann} G.,  2001,
  \mn@doi [\mnras] {10.1046/j.1365-8711.2001.04912.x}, \href
  {http://adsabs.harvard.edu/abs/2001MNRAS.328..726S} {328, 726}

\bibitem[\protect\citeauthoryear{{Springel}, {Di Matteo}  \&
  {Hernquist}}{{Springel} et~al.}{2005a}]{Springel2005a}
{Springel} V.,  {Di Matteo} T.,   {Hernquist} L.,  2005a, \mn@doi [\mnras]
  {10.1111/j.1365-2966.2005.09238.x}, \href
  {http://adsabs.harvard.edu/abs/2005MNRAS.361..776S} {361, 776}

\bibitem[\protect\citeauthoryear{{Springel}, {Di Matteo}  \&
  {Hernquist}}{{Springel} et~al.}{2005b}]{Springel2005c}
{Springel} V.,  {Di Matteo} T.,   {Hernquist} L.,  2005b, \mn@doi [\apjl]
  {10.1086/428772}, \href {http://adsabs.harvard.edu/abs/2005ApJ...620L..79S}
  {620, L79}

\bibitem[\protect\citeauthoryear{{Steinborn}, {Hirschmann}, {Dolag}, {Shankar},
  {Juneau}, {Krumpe}, {Remus}  \& {Teklu}}{{Steinborn}
  et~al.}{2018}]{Steinborn2018}
{Steinborn} L.~K.,  {Hirschmann} M.,  {Dolag} K.,  {Shankar} F.,  {Juneau} S.,
  {Krumpe} M.,  {Remus} R.-S.,   {Teklu} A.~F.,  2018, \mn@doi [\mnras]
  {10.1093/mnras/sty2288}, \href
  {https://ui.adsabs.harvard.edu/abs/2018MNRAS.481..341S} {481, 341}

\bibitem[\protect\citeauthoryear{{The EAGLE team}}{{The EAGLE
  team}}{2017}]{EAGLE2017}
{The EAGLE team} 2017, preprint, \href
  {http://adsabs.harvard.edu/abs/2017arXiv170609899T} {} (\mn@eprint {arXiv}
  {1706.09899})

\bibitem[\protect\citeauthoryear{{Trayford} et~al.,}{{Trayford}
  et~al.}{2015}]{Trayford2015}
{Trayford} J.~W.,  et~al., 2015, \mn@doi [\mnras] {10.1093/mnras/stv1461},
  \href {http://adsabs.harvard.edu/abs/2015MNRAS.452.2879T} {452, 2879}

\bibitem[\protect\citeauthoryear{{Villforth} et~al.,}{{Villforth}
  et~al.}{2014}]{Villforth2014}
{Villforth} C.,  et~al., 2014, \mn@doi [\mnras] {10.1093/mnras/stu173}, \href
  {http://adsabs.harvard.edu/abs/2014MNRAS.439.3342V} {439, 3342}

\bibitem[\protect\citeauthoryear{{Villforth} et~al.,}{{Villforth}
  et~al.}{2017}]{Villforth2017}
{Villforth} C.,  et~al., 2017, \mn@doi [\mnras] {10.1093/mnras/stw3037}, \href
  {http://adsabs.harvard.edu/abs/2017MNRAS.466..812V} {466, 812}

\bibitem[\protect\citeauthoryear{{Volonteri}, {Capelo}, {Netzer}, {Bellovary},
  {Dotti}  \& {Governato}}{{Volonteri} et~al.}{2015}]{Volonteri2015a}
{Volonteri} M.,  {Capelo} P.~R.,  {Netzer} H.,  {Bellovary} J.,  {Dotti} M.,
  {Governato} F.,  2015, \mn@doi [\mnras] {10.1093/mnras/stv387}, \href
  {http://adsabs.harvard.edu/abs/2015MNRAS.449.1470V} {449, 1470}

\bibitem[\protect\citeauthoryear{{Weston}, {McIntosh}, {Brodwin}, {Mann},
  {Cooper}, {McConnell}  \& {Nielsen}}{{Weston} et~al.}{2017}]{Weston2017}
{Weston} M.~E.,  {McIntosh} D.~H.,  {Brodwin} M.,  {Mann} J.,  {Cooper} A.,
  {McConnell} A.,   {Nielsen} J.~L.,  2017, \mn@doi [\mnras]
  {10.1093/mnras/stw2620}, \href
  {http://adsabs.harvard.edu/abs/2017MNRAS.464.3882W} {464, 3882}

\bibitem[\protect\citeauthoryear{{Wiersma}, {Schaye}  \& {Smith}}{{Wiersma}
  et~al.}{2009a}]{Wiersma2009a}
{Wiersma} R.~P.~C.,  {Schaye} J.,   {Smith} B.~D.,  2009a, \mn@doi [\mnras]
  {10.1111/j.1365-2966.2008.14191.x}, \href
  {http://adsabs.harvard.edu/abs/2009MNRAS.393...99W} {393, 99}

\bibitem[\protect\citeauthoryear{{Wiersma}, {Schaye}, {Theuns}, {Dalla Vecchia}
   \& {Tornatore}}{{Wiersma} et~al.}{2009b}]{Wiersma2009b}
{Wiersma} R.~P.~C.,  {Schaye} J.,  {Theuns} T.,  {Dalla Vecchia} C.,
  {Tornatore} L.,  2009b, \mn@doi [\mnras] {10.1111/j.1365-2966.2009.15331.x},
  \href {http://adsabs.harvard.edu/abs/2009MNRAS.399..574W} {399, 574}

\bibitem[\protect\citeauthoryear{{Zolotov} et~al.,}{{Zolotov}
  et~al.}{2015}]{Zolotov2015}
{Zolotov} A.,  et~al., 2015, \mn@doi [\mnras] {10.1093/mnras/stv740}, \href
  {http://adsabs.harvard.edu/abs/2015MNRAS.450.2327Z} {450, 2327}

\makeatother
\end{thebibliography}
\appendix

\section{Choice of parameters}
\label{sect:choice_of_parameters}

In this appendix we explore how sensitive the results of this study are to our choice of parameters, that is: the definition of a \squotes{merging state}, the definition of an \squotes{active} BH and how we match a selected galaxy to a control galaxy. We note, that in \cref{sect:define_active} we only explicitly describe the changes to the results when considering the bolometric AGN luminosity, as the differences when considering the Eddington rate are so similar. In \cref{sect:choosing_control_sample} we explicitly describe the changes when considering both the bolometric AGN luminosity and Eddington rate separately. 

\subsection{Defining a \squotes{merging state} and an \squotes{active} BH}
\label{sect:define_active}

In \cref{sect:merger_fraction} we investigated the merger fraction of AGN by assuming that galaxies were in the \squotes{state of a merger} if they have recently undergone coalescence with another galaxy up to one dynamical time in the past, or will undergo coalescence with another galaxy up to one dynamical time in the future (i.e., $a=-1$ and $b=1$ in \cref{eq:merger_fraction}). This allowed us to identity major merging systems as those with a value of \ndynmajorbar $\leq 1$, and \squotes{isolated} systems as those with a value of \ndynmajorbar $> 1$ (although for this study we used \ndynmajorbar $> 2$ to classify an isolated system to be conservative). Whilst one dynamical time has physical meaning in relation to the dynamics of a system during an interaction, it is still somewhat an arbitrary choice.

\begin{figure} \includegraphics[width=\columnwidth]{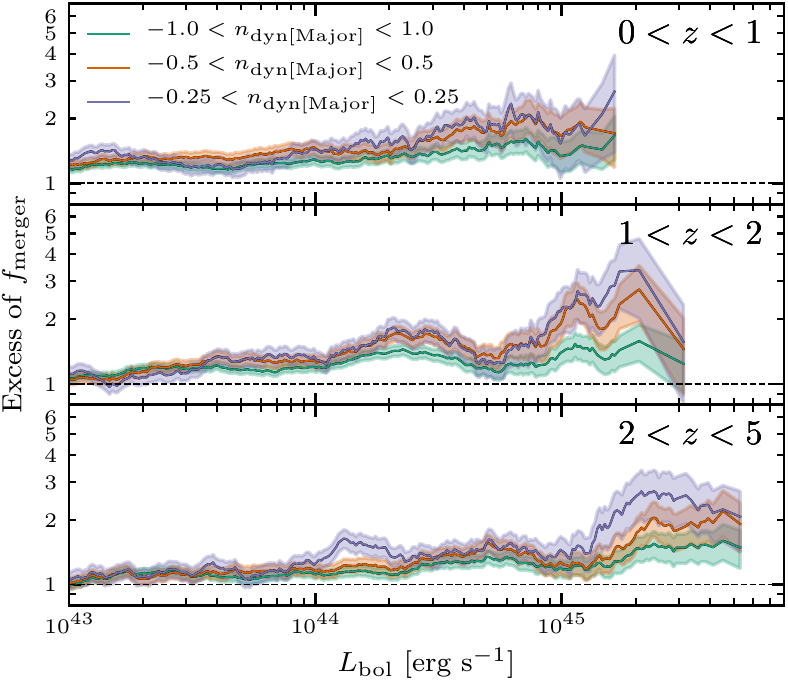}

\caption{How the excess in the major merger fraction of AGN as a function of the bolometric AGN luminosity (i.e., the results from lower left panel of \cref{fig:major_merger_fraction}) varies with how we define a \squotes{merging state} (see \cref{eq:merger_fraction}). Our fiducial value, $-1.0 < n_{\mathrm{dyn[Major]}} < 1.0$, typically produces lower excess values than if we were to consider a smaller dynamical time window, with the excess values potentially varying by up to a factor of $\approx 2$ at the highest AGN luminosities, i.e., \Lbol $\gtrsim 10^{45}$~\erg, depending on the choice of dynamical time window (albeit with large errors).}

\label{fig:time_excess_compare}
\end{figure}

In \cref{fig:time_excess_compare} we investigate how the excess of the merger fraction between galaxies with active and inactive BHs (i.e., the results from the lower left panel of \cref{fig:major_merger_fraction}) varies as we vary the definition of a merging state. We compare the results of our fiducial definition of $\pm 1$~dynamical time to two shorter dynamical time windows: $\pm 0.5$~dynamical times and $\pm 0.25$~dynamical times. Typically, the excess values are largest when considering a smaller dynamical time window, potentially by up to a factor of $\approx 2$ at brighter AGN luminosities (i.e., \Lbol $\gtrsim 10^{45}$~\erg) and higher redshifts ($z>1$). This results from the fact that the greatest enhancement of BH activity triggered via the merger process comes around or soon after the coalescence of the two galaxies has completed (i.e., close to \ndynmajor $\approx 0$, see \cref{fig:excess_vs_ndyn_broad,fig:excess_vs_ndyn}). We recognise that the regions where the increased excess is largest (i.e., at higher AGN luminosities) is also the region with the largest errors, and thus the values do still remain consistent with one another. Regardless, the overall behaviour of a rising excess with rising AGN luminosity appears to be largely independent of the choice of dynamical time window.  

In \cref{sect:agn_fraction} and \cref{sect:close_pairs} we compared the AGN fraction of merging (\ndynmajorbar $\leq 1$) and isolated galaxies (\ndynmajorbar $> 2$) as a function of stellar mass (see \cref{fig:agn_fraction}), and the 3D pair separation (see \cref{fig:close_pairs}). This required us to make a choice of cut to define what is and what isn't an \squotes{active} BH. For this study, when defined by a cut in the bolometric AGN luminosity an active BH has a value greater than \Lbol $\geq 10^{43}$~\erg, and when defined by a cut in the Eddington rate an active BH has a value greater than \edd $\geq 10^{-2}$.

\begin{figure} \includegraphics[width=\columnwidth]{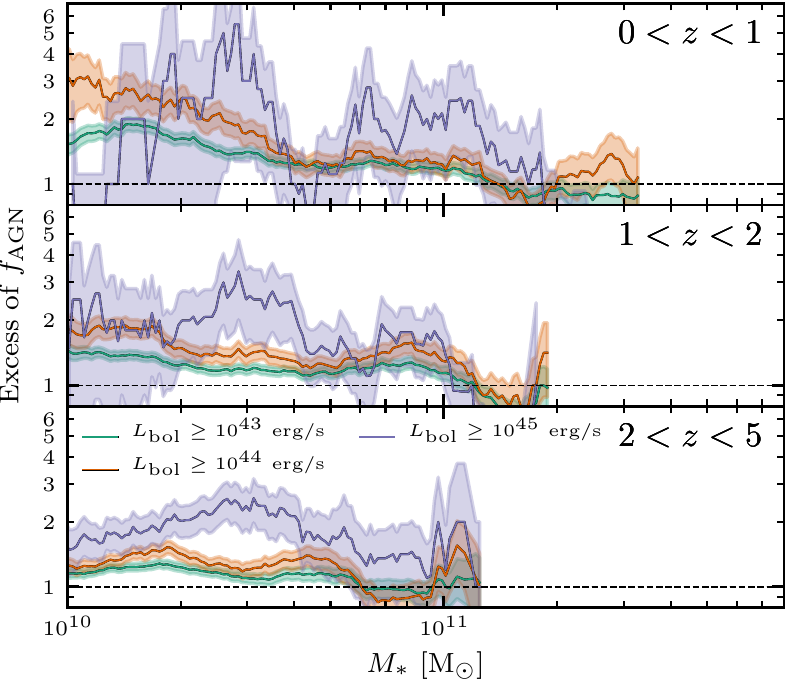}

\caption{How the excess in the AGN fraction (defined by a cut in the bolometric AGN luminosity) as a function of the stellar mass (i.e., the results from lower left panel of \cref{fig:agn_fraction}) varies with how we define an \squotes{active} BH. Our fiducial cut, \Lbol $\geq 10^{43}$~\erg, typically produces lower excess values than if we were to consider a higher cut in the bolometric luminosity, with the excess values potentially varying by up to a factor of $\approx 3$--4 at lower redshifts depending on the choice of cut (albeit with large errors).}

\label{fig:agn_excess_compare}
\end{figure}

In \cref{fig:agn_excess_compare} we test how the choice of bolometric AGN luminosity cut affects the excess in the AGN fraction between major merging and isolated systems (i.e., the results from the lower left panel of \cref{fig:agn_fraction}). We compare our fiducial cut of \Lbol $\geq 10^{43}$~\erg to two higher luminosity cuts: \Lbol $\geq 10^{44}$~\erg and \Lbol $\geq 10^{45}$~\erg. Typically, higher cuts in the AGN luminosity result in higher excess values: increasing by up to a factor of $\approx 2$ at higher redshifts ($1<z<5$) and potentially increasing by up to a factor of $\approx 3$--4 at lower redshifts ($0<z<1$, albeit with large errors). At redshifts below $z<2$, the trends of an increasing excess in the AGN fraction with decreasing stellar mass are also much more pronounced at the highest AGN luminosity cut we explore (\Lbol $\geq 10^{45}$~\erg), however the overall behaviour is largely similar regardless of the luminosity cut. These results suggest that the excess values are potentially quite sensitive to the choice of AGN cut.  

\begin{figure} \includegraphics[width=\columnwidth]{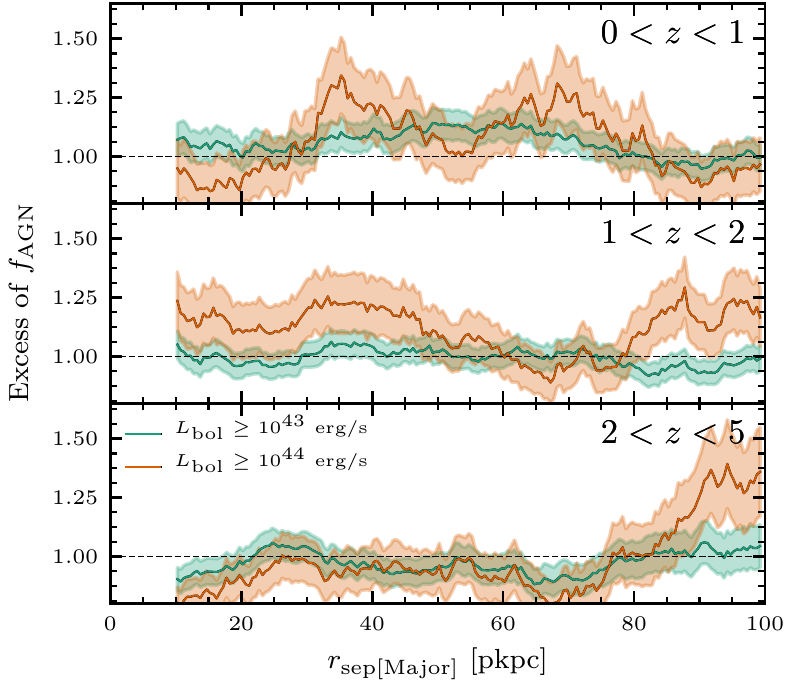}

\caption{How the excess in the AGN fraction (defined by a cut in the bolometric AGN luminosity) as a function of the 3D pair separation between the two galaxies (i.e., the results from the lower left panel of \cref{fig:close_pairs}) varies with how we define an \squotes{active} BH. Our fiducial cut, \Lbol $\geq 10^{43}$~\erg, typically produces lower excess values than if we were to consider a higher cut in the bolometric luminosity.}

\label{fig:agn_excess_compare_close_pairs}
\end{figure}

Similarly, in \cref{fig:agn_excess_compare_close_pairs} we test how the choice of bolometric AGN luminosity cut affects the excess in the AGN fraction between galaxies with close major companions and isolated galaxies (i.e., the results from the lower left panel of \cref{fig:close_pairs}). We compare our fiducial cut of \Lbol $\geq 10^{43}$~\erg to a higher luminosity cut of \Lbol $\geq 10^{44}$~\erg (luminosity cuts any higher than this have too few numbers to adequately explore within the simulation). Similar to \cref{fig:agn_excess_compare}, we find the greatest excess in the AGN fractions above the isolated control galaxies come with higher luminosity cuts (at least for redshifts $z<2$). 

Therefore the choice of how we define a \squotes{merging state} and \squotes{active} BH does impact the results, and therefore needs to be considered when comparing to similar studies of this nature. It is also for this reason why one should be careful when comparing the merger and AGN fractions, and the resulting excess values, between the predictions of the simulation and the observations.   

\subsection{The choice of parameters to match a selected galaxy to a control galaxy for forming a control sample}
\label{sect:choosing_control_sample}

%\begin{figure} \includegraphics[width=\columnwidth]{plots/ControlCheck_merger_fraction_edd_Major_default.pdf}

%\caption{The median properties of the galaxies within Eddington rate selected sample (solid lines) compared to the median properties of the associated inactive control galaxies (dashed lines). Here, the control galaxies are matched using our \squotes{intermediate} criteria (i.e., they are matched on $M_*$, $z$, $N_2$ and $r_2$, resulting in the two samples agreeing closely on their values for these properties). By matching on these four parameters, this naturally selects control galaxies residing in haloes of very similar masses. However, as the AGN with high Eddington rates are gas-rich for galaxies of their stellar masses, and have undermassive BHs for galaxies of their stellar masses, the control galaxies are systematically higher in their BH masses and lower in their gas masses than the selected galaxies. This bias between the selected high Eddington rate AGN and their control galaxies is why the resulting excess in the merger fractions seen in \cref{fig:excess_compare} can vary by over a factor of $\approx 3$ depending on the matching criteria used.}

%\label{fig:agn_vs_control}
%\end{figure}

\begin{figure*} \includegraphics[width=\textwidth]{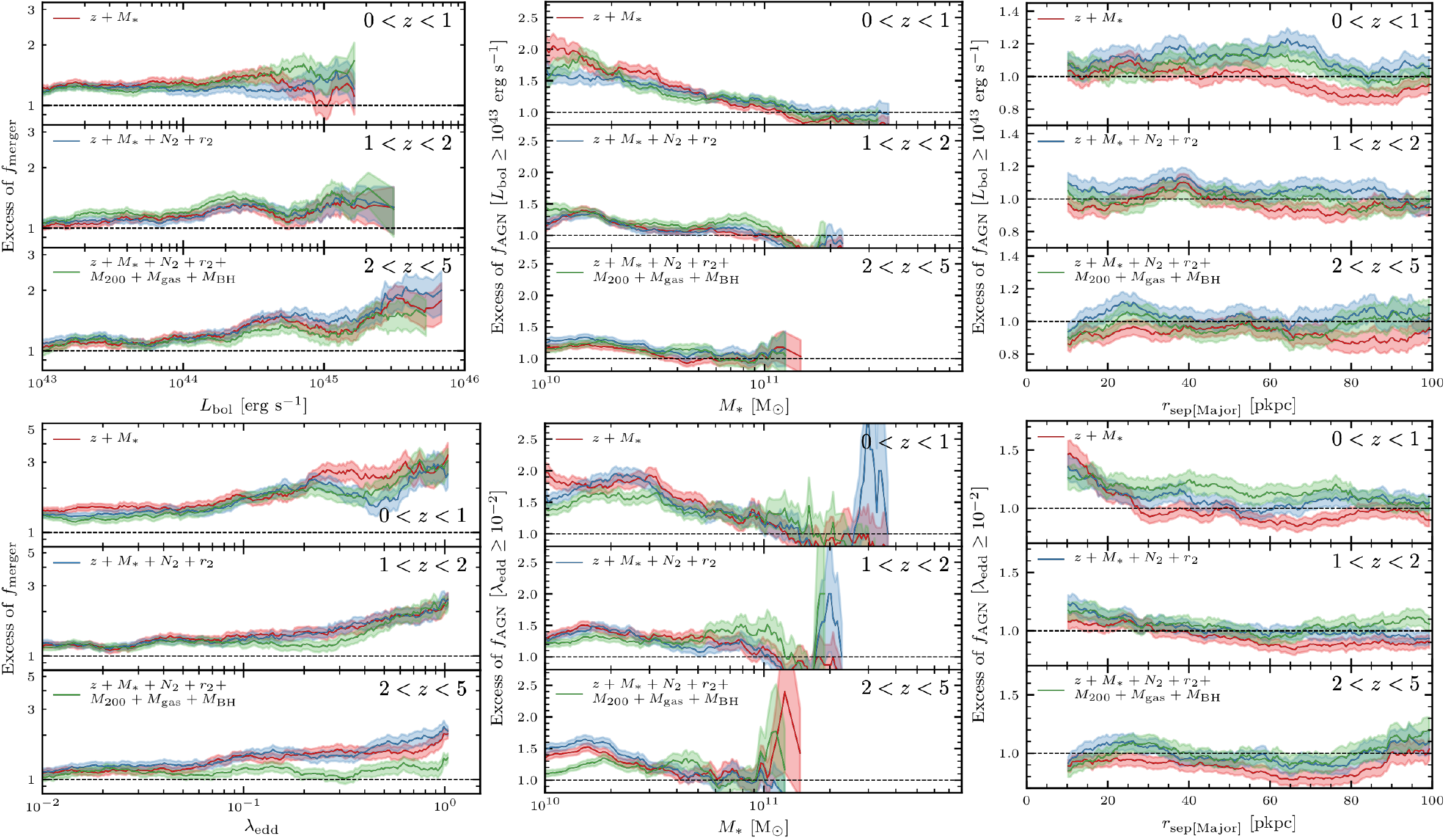}

\caption{How the excess in the major merger fractions from \cref{fig:major_merger_fraction} (left two panels), the excess in the AGN fractions from \cref{fig:agn_fraction} (middle two panels), and the excess in the AGN fractions from \cref{fig:close_pairs} (right two panels) change depending on the control galaxy matching criteria that is used. The legend shows what properties are matched between the selected galaxies and their associated control galaxies. Broadly speaking, the values of the excess are higher when fewer parameters are matched. However, whilst the values of the excess can change depending on the matching criteria used, the overall behaviour in each panel is largely unaffected by the choice of matching criteria.} 
 
\label{fig:excess_compare}
\end{figure*}

Throughout this study we have investigated to what extent galaxy--galaxy mergers enhance BH activity, which we've chosen to quantify by a fractional \squotes{excess} in BH activity relative to a control sample. For \cref{sect:merger_fraction} it was the excess between the merger fraction of AGN (\Lbol $\geq 10^{43}$~\erg or \edd $\geq 10^{-2}$) and a control sample of inactive galaxies (\Lbol $< 10^{43}$~\erg or \edd $< 10^{-2}$), for \cref{sect:agn_fraction} it was the excess between the AGN fraction of merging systems (\ndynmajorbar $\leq 1$) and a control sample of isolated galaxies (\ndynmajorbar $>2$), and for \cref{sect:close_pairs} it was the excess between the AGN fraction of galaxies with major close companions ($r_{\mathrm{sep[Major]}} \leq 100$~pkpc) and a control sample of isolated galaxies ($r_{\mathrm{sep[Major]}} > 200$~pkpc). The interpretation of our results, therefore, is sensitive to the value of this excess, which is sensitive to how the galaxies within the selected sample are matched to a control galaxy counterpart (see \cref{sect:control_sample}). Here we investigate to what extent the matching criteria by which we choose our control galaxies impacts our results.

It has been well established that: (1) the merger fraction of galaxies at fixed mass increases with increasing redshift, and (2) the merger fraction of galaxies at fixed redshift increases with increasing mass \citep[e.g.,][]{Rodighiero2015,Qu2017}. It is therefore essential that any paired control galaxy must at least match on the stellar mass and redshift. This two-part criteria is how many observational studies of this nature have selected their control galaxies, as it is often all that can be feasibly achieved. Some observational studies have extended this minimalist criteria by also considering the role the environment, by additionally matching the control galaxies on the $N_2$ and $r_2$ parameters \citep[e.g.,][]{Patton2013,Patton2016}. For this study we wanted to ensure that the control galaxies were as similar as possible to the selected galaxies, opting for a criteria that matches on the stellar, gas, BH and halo masses and also on the $N_2$ and $r_2$ parameters. 

To test their impact, here we experiment with three matching criteria:

\begin{enumerate}
    \item A \squotes{basic} criteria, matching only on the stellar mass and redshift ($M_* + z$).
    \item An \squotes{intermediate} criteria, which additionally matches on the environment ($M_* + z + N_2 + r_2$).
    \item A \squotes{strict} criteria, which further matches on the BH mass, gas mass and halo mass ($M_* + z + N_2 + r_2 + M_{\mathrm{BH}} + M_{\mathrm{gas}} + M_{\mathrm{200}}$).
\end{enumerate}

\noindent We note that we have deliberately chosen to avoid matching on the SFR, as the SFR of a galaxy can also be enhanced during the merger process.

To see how the choice of matching criteria impacts the results of this study, we include \cref{fig:excess_compare}. This shows the excess merger and AGN fractions from the lower panels of \cref{fig:major_merger_fraction,fig:agn_fraction,fig:close_pairs}, now repeating the analysis for each of the three matching criteria. We find that when the bolometric AGN luminosity is considered, the control pairing criteria has only a slight overall impact on the measured excess. The measured excess is slightly more sensitive to the control pairing criteria when the Eddington rate is considered, however the values never deviate from one another by more than a factor of two, and their errors are often overlapping. Typically, when fewer parameters are matched, the \emph{higher} the values of the excess. However, regardless of the matching criteria used, the behaviour of the trends is unchanged. Therefore whilst the excess values do change with the choice of matching criteria, the interpretation of the results is unaffected.

% Don't change these lines
\bsp	% typesetting comment
\label{lastpage}

\end{document}